\documentclass[useAMS,usenatbib,usegraphicx,onecolumn]{mn2e}

\usepackage{natbib,aas_macros,color,comment,ulem}

\title[Connecting the Sun and the Solar Wind: The Self-consistent Transition of Heating Mechanisms]{Connecting the Sun and the Solar Wind: The Self-consistent Transition of Heating Mechanisms}
\author[T. Matsumoto and T. K. Suzuki]
{T. Matsumoto\thanks{E-mail:takuma.matsumoto@nagoya-u.jp (TM);stakeru@nagoya-u.jp (TKS)} and 
 T. K. Suzuki\footnotemark[1]  \\
 Department of Physics, Nagoya University, Furo-cho, Chikusa-ku, Nagoya, Aichi 464-8602, Japan}
\begin{document}

\date{Accepted 1988 December 15. Received 1988 December 14; in original form 1988 October 11}

\pagerange{\pageref{firstpage}--\pageref{lastpage}} \pubyear{2002}

\maketitle

\label{firstpage}

\begin{abstract}
We have performed a 2.5 dimensional magnetohydrodynamic simulation that resolves the propagation and dissipation of Alfv\'{e}n waves in the solar atmosphere. 
Alfv\'{e}nic fluctuations are introduced on the bottom boundary of the extremely large simulation box that ranges 
from the photosphere to far above the solar wind acceleration region.
Our model is ab initio in the sense that no corona and no wind are assumed initially.
The numerical experiment reveals the quasi-steady solution that has the transition from the cool to the hot atmosphere and the emergence of the high speed wind.
The global structure of the resulting hot wind solution fairly well agree with the coronal and the solar wind structure inferred from observations.
The purpose of this study is to complement the previous paper by \cite{2012ApJ...749....8M} and describe the more detailed results and the analysis method.
These results include the dynamics of the transition region and the more precisely measured heating rate in the atmosphere.
Particularly, the spatial distribution of the heating rate helps us to interpret the actual heating mechanisms in the numerical simulation.
Our estimation method of heating rate turned out to be a good measure for dissipation of 
Alfv\'{e}n waves and low beta fast waves.
\end{abstract}


\section{Introduction}

For years, extensive studies have been carried out on the coronal heating problem.
The hot corona lies above the cool photosphere \citep{1943ZA.....22...30E}, which can not be explained by thermal processes.
Instead the mechanical energy injection from the surface convection motion 
is expected to maintain the hot corona by means of magnetic field.
The coupling between the convection and the magnetic field transports free magnetic energy that will be converted to thermal energy
in the upper atmosphere.
To resolve the coronal heating problem, we should specify the physics that are responsible for releasing the free magnetic energy.

In this paper, we will focus on the atmosphere above the solar pole where the fast solar wind emanates.
The magnetic field lines above the solar pole open up to the interplanetary space.
One of the most plausible carriers of the magnetic free energy in the open field region is Alfv\'{e}n waves,
which was first proposed by \cite{1947MNRAS.107..211A}.
Recent observations have been accumulating several pieces of evidence of the existence of 
the propagating Alfv\'{e}n waves in the solar atmosphere \citep{2011ApJ...736L..24O}.
Propagation and dissipation of Alfv\'{e}n waves in the solar atmosphere is quite complicated since
the gravity and the magnetic field make highly inhomogeneous atmospheres.
For 1 dimensional configuration, nonlinear steepening of Alfv\'{e}n waves and the subsequent shock formation 
play an important role in wave dissipation \citep{1982SoPh...75...35H,1999ApJ...514..493K,2010ApJ...710.1857M}.
Plenty of wave modes will emerge when we consider two dimensional configuration \citep{1983PASJ...35..263S,2002ApJ...564..508R,2003ApJ...599..626B}.

All the above numerical simulations adopt simplified energy equation.
In the coronal loop, mechanical heating is balanced by the radiative cooling and thermal conduction.
Enthalpy flux due to the existence of the solar wind is also important when we consider the open field structure \citep{1986JGR....91.4111H}.
Therefore in order to obtain the atmospheric structure above the solar pole, we should treat the corona 
and the solar wind in a self-consistent way.
One of the most important aspects of the self-consistent models is that these models can determine the mass loss rate
from the sun given the boundary condition at the surface.
If we can construct the appropriate mass loss model from the sun, the model could be very useful even for the 
stellar or the planetary atmospheres.

Pioneering works by \cite{1982ApJ...259..779H,1982ApJ...259..767H} treated 
the corona and the solar wind simultaneously in a self-consistent way.
Instead of using the artificial heating function, \cite{1986JGR....91.4111H,2007ApJS..171..520C} adopted 
the physically based heating model from phenomenological turbulent theories.
Heating models based on acoustic shocks \citep{2002ApJ...578..598S} and MHD shocks \citep{2004MNRAS.349.1227S}
are also investigated.
Time steady condition is relaxed 
to demonstrate the self-consistent reproduction of the hot coronal wind with fully dynamical 
1D \citep{2005ApJ...632L..49S,2006JGRA..11106101S} and 2D \citep{2012ApJ...749....8M} (hereafter referred to as MS12) 
MHD simulations.

MS12 suggested that the shock heating is dominant heating mechanisms in the chromosphere and the coronal bottom
while the turbulent heating is important in the solar wind acceleration region.
However, there are two problems in the way how MS12 estimate the heating rate in their simulation.
First, MS12 used dimensional analysis based on turbulent theory when they estimate the incompressible heating rate.
This method can include large uncertainty, which allows MS12 only to estimate the order of magnitude of the heating rate.
Second, the dimensional analysis does not have ability to estimate 
the temporal and spatial variability of the heating rate.
In order to overcome the problems in MS12, we have developed an alternative method to estimate the heating rate.
The purpose of the present paper is to describe more detailed features on heating mechanisms by using the new estimation method.
Moreover, we will describe detailed dynamic features which are not shown in MS12.

We begin in section \ref{sec:model} by describing our numerical models and corresponding assumptions.
The time-averaged quasi-steady solution is shown in section \ref{sec:solution} while the dynamic features 
relating to the coupling between waves and the transition region shall be discussed in section \ref{sec:dynamics}.
Then the basic idea to derive the heating rate as a function of space and time is given in section 
\ref{sec:data_analysis}, although the detailed method for discretization and test simulations 
are explained in appendix \ref{app:method}.
In section \ref{sec:heating}, we shall show the total amount and the spatial distribution of heating rate and 
discuss the possible mechanisms that actually happened in the numerical simulation.
We continue in section \ref{sec:discussion} by discussing the validity of the heating mechanisms 
found in our numerical simulation, while in section \ref{sec:summary} we summarize.

\section[]{Model \& Assumptions}\label{sec:model}

\begin{figure}
      \begin{center}
      \includegraphics[scale=1.0]{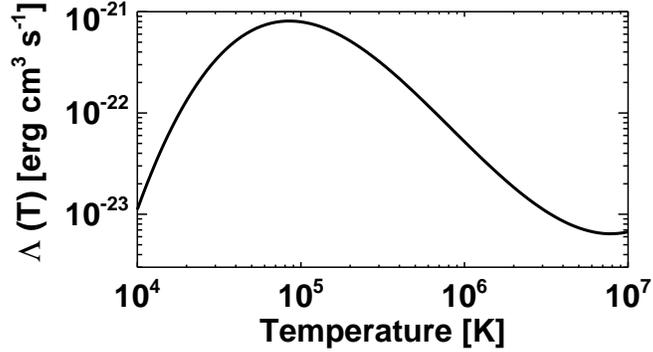}
      \caption{ The radiative loss function as a function of temperature.
              }
      \label{fig:lambda_func}
      \end{center}
   \end{figure}

   All the numerical methods to integrate the basic equations in this paper are the same as that used in MS12.
   We will describe the method in more detailed way here.
   Throughout the paper, we will assume single fluid MHD to describe a flux tube extended 
   from the photosphere to the interplanetary space.
   Although the photosphere and the chromosphere are partially ionized, 
   the use of single fluid approximation can be justified by frequent collisions between 
   protons and neutrals \citep[e.g.,][]{2008MNRAS.385.2269P}.
   Above the upper corona ($r-r_{\rm s} > r_{\rm s}$ or 7.0$\times$10$^2$ Mm), 
   proton and electron will decouple due to weak collisionality.
   Even in that case, we continue to use single fluid MHD equations for simplicity.
   Since we do not include any explicit dissipation terms in our equations,
   all the dissipations come from shocks and discretization errors.
   The method to estimate the heating rate shall be discussed in section \ref{sec:data_analysis}.
   In addition to ideal MHD formulation, we include the gravity from the Sun.
   The solar gravity produces a highly stratified atmosphere 
   , which inevitably leads initially linear waves to nonlinear waves.
   We consider a flux tube at $\theta=\pi/2$ in the spherical coordinate.
   We rotate the axis of the spherical coordinate so that the pole of the spherical coordinate 
   lies on the equator of the sun. This helps us to avoid the numerical difficulty in the polar region
   of the spherical coordinate when we treat polar region of the sun.
   We neglect rotation of the sun as well as macro-scale magnetic field in this study.
   We impose the translational symmetry in the $\theta$ direction ($\partial _\theta = 0$) so that
   all the variables depend on radius ($r$), azimuthal angle ($\phi$), and time ($t$).
   Perturbations in $\theta$ direction is purely incompressible, namely Alfv\'{e}n mode, and 
   perturbations of $r$ \& $\phi$ components consist of fast and slow MHD modes (see section \ref{sec:dynamics}).
   Then our basic equations can be described as follows.

   \begin{eqnarray}
     {\partial {\cal U} \over \partial t} + {\partial {\cal F} \over \partial r}
     + {\partial {\cal G} \over \partial \phi} = {\cal S},
   \end{eqnarray}

   \begin{eqnarray}
     {\cal U} = \left(
     \begin{array}{c}
       r^2 \rho \\
       r^2 \rho V_r \\
       r^2 \rho V_\theta \\
       r^2 \rho V_\phi \\
       r^2 B_r \\
       rB_\theta \\
       rB_\phi \\
       r^2 E 
     \end{array}     \right) \label{eq:cnsvar}
     \equiv \left(
     \begin{array}{c}
       {\cal R}\\
       {\cal M}_r\\
       {\cal M}_\theta\\
       {\cal M}_\phi\\
       {\cal B}_r\\
       {\cal B}_\theta\\
       {\cal B}_\phi\\
       {\cal E}
     \end{array}     \right) ,
   \end{eqnarray}

   \begin{eqnarray}
     {\cal S} = \left(
     \begin{array}{c}
       0 \\
       \left[ - B_\perp^2 + \rho V_\perp^2 + 2 P_{\rm T} - {\rho GM/r} \right] \times r \\
       \left( B_{r} B_\theta - \rho V_{r} V_\theta \right) \times r\\
       \left( B_{r} B_\phi - \rho V_{r} V_\phi \right) \times r\\
       0 \\
       0 \\
       0 \\
       -\rho V_{r} G M + Q_{\rm cnd} + Q_{\rm rad}
     \end{array}     \right) ,
   \end{eqnarray}

   \begin{eqnarray}
     {\cal F} = 
     \left(
     \begin{array}{c}
       \rho V_{r} \times r^2 \\
       \left( \rho V_{r}^2 + P_{\rm T} - B_{r}^2\right) \times r^2 \\
       \left( \rho V_\theta V_{r}  - B_{\theta} B_{r} \right) \times r^2 \\
       \left( \rho V_\phi V_{r}  - B_{\phi} B_{r} \right) \times r^2 \\
       0 \\
       \left( V_{r} B_\theta - V_\theta B_{r} \right) \times r \\
       \left( V_{r} B_\phi - V_\phi B_{r} \right) \times r \\
       \left[ \left( E + P_{\rm T} \right) V_{r} 
       - \left( \bmath{V} \cdot \bmath{B} \right) B_{r} \right] \times r^2
     \end{array}     \right) ,
   \end{eqnarray}
   \begin{eqnarray}
     {\cal G} = \left(
     \begin{array}{c}
       \rho V_\phi \times r \\
       \left( \rho V_{r} V_\phi - B_{r} B_\phi \right) \times r \\
       \left( \rho V_\theta V_\phi - B_\theta B_\phi \right) \times r \\
       \left( \rho V_\phi^2 + P_{\rm T} - B_\phi^2\right) \times r \\
       \left( V_\phi B_{r} - V_{r} B_\phi \right) \times r \\
       V_\phi B_\theta - V_\theta B_\phi \\
       0 \\
       \left[ \left( E + P_{\rm T} \right) V_\phi 
       - \left( \bmath{V} \cdot \bmath{B} \right) B_\phi \right] \times r
     \end{array}     \right) ,
   \end{eqnarray}

   \begin{eqnarray}
     P_{\rm g} &=& {\rho k_{\rm B} T \over \mu m_{\rm H}},\\
     E &=& {1\over2}\rho V^2 +{P_{\rm g}\over (\gamma-1)} + {B^2\over 2}, \\
     P_{\rm T} &=& P_{\rm g} + {B^2 \over 2},\\
     V_{\perp}^2 &=& V_\theta^2 + V_\phi^2, \\
     B_{\perp}^2 &=& B_\theta^2 + B_\phi^2,
   \end{eqnarray}
   where all the symbols have standard meanings 
   except for magnetic field, $B$, which is normalized by $\sqrt{4 \pi}$.
   We adopt the specific heat ratio $\gamma$ of $5/3$ for mono-atomic molecules.
   At around 6,000 K or less, almost all of hydrogens are neutrals while they are 
   completely ionized at the temperature above 10$^4$ K.
   To mimic the real equation of states there, we define the mean molecular weight 
   as a function of temperature to linearly connect fully ionized and neutral gases,
   \begin{eqnarray}
     \mu = \left\{
     \begin{array}{c}
       1 ~~\left( T < 6,000 ~{\rm K} \right) \\
       \\
       \left[ 3.5 \left( 1-4285.7/T \right) \right]^{-1} \\
       \left( 6,000 ~{\rm K} < T < 10^4 ~{\rm K} \right) \\
       \\
       0.5 ~~\left( {\rm otherwise} \right)
     \end{array} \right. .
   \end{eqnarray}
   As for the thermal conduction, we use the classical formulation for collisional plasma, 
   \begin{equation}
     Q_{\rm cnd} = \nabla \cdot \left( \kappa_0 T^{5/2} 
     { \nabla T \cdot \bmath{B} \over B^2} \bmath{B} \right).
     \label{qcnd}
   \end{equation}
   We choose $\kappa_0 = 10^{-6}$ in cgs unit, reasonable value of electrons for fully ionized plasma 
   in thermal equilibrium \citep{1965RvPP....1..205B}.
   Heat transport across magnetic field lines is inhibited in equation (\ref{qcnd}).
   In the present study, we use this classical limit throughout all space, although 
   the deviation from the the classical heat conduction could affect the temperature structure in the outer corona 
   ($>10$ $r_{\rm s}$ or 7.0 $\times$ 10$^3$ Mm)
   \citep[see][]{1978RvGSP..16..689H}.
   We also introduce radiative cooling as follows.
   \begin{eqnarray}
     Q_{\rm rad} = \left\{
     \begin{array}{lc}
       - n^2 \Lambda \left( T \right) & ~(T>T_{\rm c} ~\&~ \rho< \rho _{\rm c})\\
       - n n_{\rm c} \Lambda \left( T_{\rm c} \right) &({\rm otherwise}) \label{eq:radloss}
     \end{array} \right. ,
     \label{qrad}
   \end{eqnarray}
   \begin{eqnarray}
     \begin{array}{cllc}
       \log _{10} \Lambda \left( T \right) &=-&127.3 & \\
       &+& 56.57 & \log _{10} T  \\
       &-& 9.84 & (\log _{10} T )^2 \\
       &+& 0.5548 & (\log _{10} T )^3 \label{eq:lambda_func},
     \end{array}
   \end{eqnarray}
   where $n=\rho/m_{\rm H}$, $T_{\rm c} = 4 \times 10^4$ K, and $\rho_{\rm c} = 4.9 \times 10^{-17}$ g cm$^{-3}$.
   The term in the first line in equation (\ref{eq:radloss}) represents the radiative cooling for optically thin plasma of the coronal abundance 
   \citep[e.g.,][]{1990A&AS...82..229L} fitted by polynomials.
   The radiative loss function is shown in figure \ref{fig:lambda_func}.
   The term in the second line of equation (\ref{eq:radloss}) is in proportion to $n$ and mimics the radiative cooling for optically thick plasma 
   that was empirically derived by \cite{1989ApJ...346.1010A}.
   Basically, the treatment of the thermal conduction and the radiative cooling in our study is the same one
   as that used in \cite{2004ApJ...601L.107M,2005ApJ...632L..49S,2006JGRA..11106101S,2008ApJ...688..669A,
   2010ApJ...712..494A}.

\begin{figure}
      \begin{center}
      \includegraphics[scale=1.0]{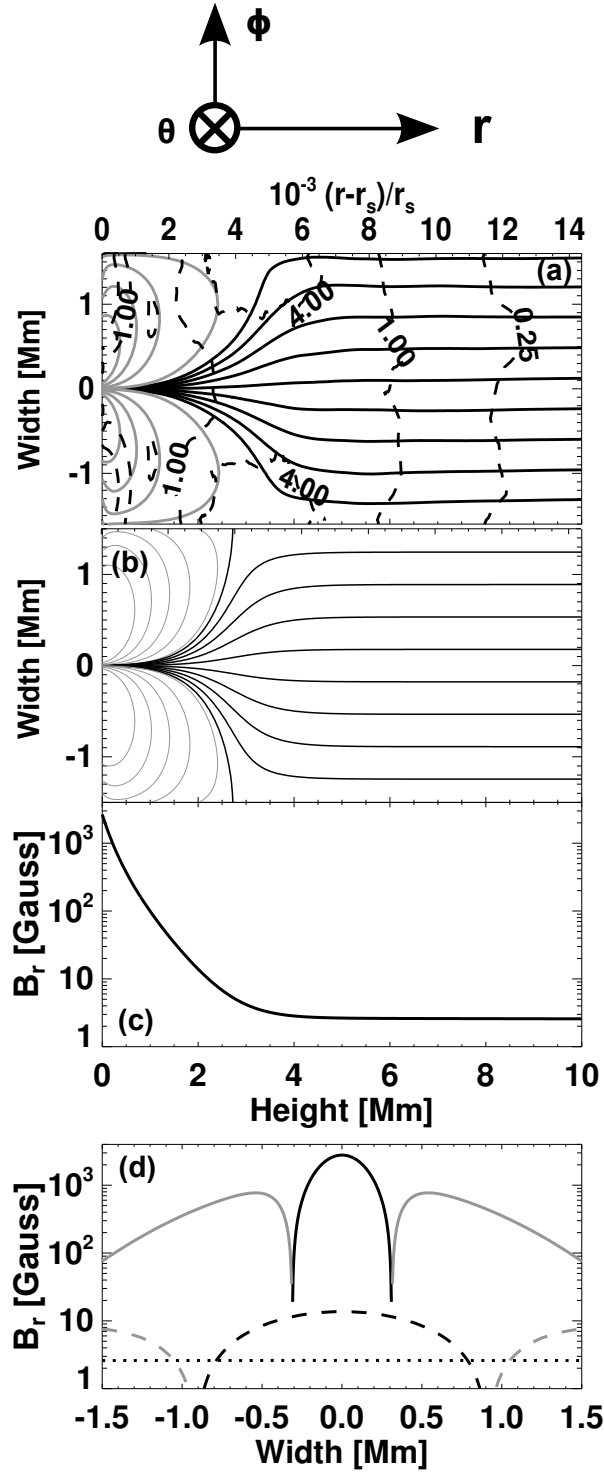}
         \caption{Magnetic field structure in our numerical simulation.
                  (a) Mean magnetic field structure in the quasi-steady state below 10 Mm (1.4$\times$10$^{-2}$ 
                  $r_{\rm s}$).
                  The black(gray) solid lines show open(closed) magnetic field lines.
                  The black dashed lines indicate iso contours of plasma beta.The magnetic field lines 
                  (b) The initial magnetic field lines below 10 Mm (1.4$\times$10$^{-2}$ $r_{\rm s}$).
                  The black(gray) solid lines represent the open(closed) field lines.
                  (c) The radial magnetic field strength along $\phi = 0$ as a function of height.
                  (d) The radial magnetic field strength along the azimuthal direction.
                  The solid, dashed, and dotted lines represent the field strength at $r-r_{\rm s}=0,2,6$ Mm
                  (0, 2.9$\times$10$^{-3}$, 8.6$\times$10$^{-3}$ $r_{\rm s}$), respectively.
                  The black/gray color means positive/negative value.
                  }
         \label{initial_field}
      \end{center}
   \end{figure}

   We set up our initial atmosphere by solving the equations of hydrostatic equilibrium without magnetic field.
   The initial atmospheres have 10$^4$ K all over the numerical domain, starting from the initial photospheric density of
   10$^{-7}$ g cm$^{-3}$ at $r=r_{\rm s}$.
   Above the height of 11 Mm ($\sim$ 1.6 $\times10^{-2}~r_{\rm s}$), 
   the density profiles are forced to be proportional to $r^{-2}$.
   This is simply because the exponential density drop causes numerical difficulties in the higher atmosphere.
   At the photospheric surface, we impose the locally concentrated magnetic field that would correspond to 
   the kilo Gauss patch at the photosphere.
   The exact profile along $\phi$ direction at the photosphere are shown with the solid line 
   in Figure \ref{initial_field}d.
   This initial profile is produced by the superposition of narrow positive Gaussian and wide negative Gaussian.
   The presence of the kilo Gauss patches is clearly confirmed by {\it HINODE}/SOT 
   even in the polar coronal hole where the high speed solar wind originates from 
   \citep{2008ApJ...688.1374T,2012ApJ...753..157S}.
   The magnetic field above the photosphere is extrapolated by using potential field approximation 
   (Figure \ref{initial_field}b).
   The super-radial expansion of the flux tube reduces the field strength from $\sim$ 3$\times$10$^3$ to 3 Gauss 
   in the region below 4 Mm (Figure \ref{initial_field}cd).
   Above $\approx$ 4 Mm (6$\times$10$^{-3}$ $r_{\rm s}$), the field strength decreases in proportion to $r^{-2}$.
   Since all the open field lines start from a single cell in the current resolution, 
   we have performed test simulations to check the resolution dependence in appendix \ref{sec:res_test}.
   We found that as far as the energy flux at the corona is concerned, the results would not change drastically.
   The field strength 3 Gauss well below the TR could be weak compared with the usual polar field strength of 
   $\sim$ 10 Gauss.
   This is just a computational issue.
   Since we solve thermal conduction implicitly, the time step is usually determined by the Alfv\'{e}n 
   crossing time of the radial grid around the transition region ($\sim 0.1$ sec in the current resolution).
   If we reduce the field strength, we can use a longer time step restricted by the CFL condition.
   Larger field strength will produce larger Poynting flux above, which can be potential source to
   heat the corona.
   
   The horizontal length of our numerical domain is 3,000 km at the photosphere 
   and it expands in proportion to $r$.
   32 grid points are allocated to resolve 3,000 km so that 
   the horizontal spatial resolution is nearly 100 km at the photosphere.
   Periodic boundary conditions are posed on the horizontal boundaries.
   As for the radial direction, we start our simulations from the photosphere ($r=r_{\rm s}$ or 7$\times$10$^2$ Mm) to 
   the interplanetary space ($r=1,674~r_{\rm s}$ $> 7$AU or 10$^6$ Mm).
   The radial spatial resolution is 25 km at the photosphere.
   We increase the grid size non-uniformly 
     ( 0.7 $r_{\rm s}$ or 490 Mm at 7 AU)
   as we go up higher to cover the whole numerical domain 
   by using 16,384 grid points.
   The total number of the radial grid points are doubled from the previous simulation by MS12
   to better resolve the wave propagation in the low corona.
   Open boundary in radial direction could be implemented by using characteristic method, 
   although it is not straightforward to use characteristic method with thermal conduction.
   Instead, we take lengthy radial domain to avoid the numerical reflection from the top boundary and
   pose zero-derivative boundary condition there.
   At the final time of our simulation, the thermal conduction front reaches 4 AU (8.6$\times$10$^2$ r$_{\rm s}$
   or 6$\times$10$^5$ Mm) while the solar wind reaches 1 AU (2.2$\times$10$^2$ r$_{\rm s}$ or 1.5$\times$10$^5$ Mm).
   Therefore, any physical information propagating from the inner boundary can not reach the top boundary within
   the duration we considered.
   In this paper we focus on the wind structure in $r<20$ $r_{\rm s}$ ($\sim 0.1$ AU, 1.4$\times$10$^4$ Mm), 
   which is in the quasi-steady state 
   after sufficient Alfv\'{e}n crossing time.

   At the inner boundary, all the variables are fixed to the initial value except for the velocity in $\theta$ direction ($V_\theta$).
   The spectra of $V_\theta$ at the bottom boundary are prescribed to excite Alfv\'{e}n waves.
   Throughout this paper, we will restrict our self to investigate the Alfv\'{e}n mode (fluctuations in $\theta$ direction).
   The one-sided power spectrum of $V_\theta$ are defined by
   \begin{eqnarray}
     P_\nu = 2 \lim_{{\tau} \to \infty } {\langle \hat{V_\theta}(\nu)\hat{V_\theta}^*(\nu) \rangle \over {\tau} },
   \end{eqnarray}
   where $\langle \rangle$ here means temporal average over period $\tau$, 
   $\nu$ is frequency, 
   and $\hat{V_\theta}(\nu)$ are complex Fourier components of $V_\theta (t)$, 
   \begin{eqnarray}
     \hat{V_\theta}(\nu) &=& \int^{\infty}_{-\infty} V_\theta (t) e^{-2\pi i \nu t} dt.
   \end{eqnarray}
   Using the one-sided power spectrum, the total power of $V_\theta$ can be described by
   \begin{eqnarray}
     \langle V_\theta (t)^2 \rangle &=& \int^{\infty}_{0} P_\nu d\nu .
   \end{eqnarray}
   We will concentrate on the white noise case ($P_\nu \propto \nu^0$) with the total power of (2.2 km s$^{-1}$)$^2$.
   This total power seems twice larger than the observed photospheric velocity 
   \citep{2010ApJ...716L..19M}.
   Since we fixed $B_{\theta}$ to be zero at the bottom boundary, out going Els\"{a}sser variables,
   \begin{eqnarray}
     z^-_{\theta}\equiv {1\over 2} \left( V_\theta - {B_\theta \over \sqrt{\rho}} \right),
   \end{eqnarray}
   becomes the half of $V_\theta$ specified above.
   Therefore, we adopt the total power of (2.2 km s$^{-1}$)$^2$ so that 
   $\langle(z^-_\theta)^2 \rangle=$(1.1 km s$^{-1}$)$^2$.
   The wave energy flux with this boundary condition ($\rho\sim10^{-7}$ g cm$^{-3}$, $z^-_{\theta}\sim1$ km s$^{-1}$, 
   and V$_{\rm A} \sim 10$ km s$^{-1}$) becomes $\sim$10$^9$ erg cm$^{-2}$ s$^{-1}$ at the photosphere, which ensures 
   adequate energy supply provided dissipation works.
   The frequency of the fluctuations is restricted 
   to range from 2.5 $\times$ 10$^{-4}$ Hz (4,000 sec) to 2 $\times$ 10$^{-2}$ Hz (50 sec) in this study.
   The waves are specified over the entire lower boundary.
   
   Our numerical code is basically based on HLLD scheme, an approximate Riemann solver that have 
   robustness and inexpensive numerical cost \citep{2005JCoPh.208..315M}.
   We combined HLLD scheme and flux-CT method \citep{2000JCoPh.161..605T,2005JCoPh.205..509G} to preserve
   the initial $\nabla \cdot \bmath{B}$ within the rounding error.
   The initial $\nabla \cdot \bmath{B}$ can be as small as the rounding error if we use 
   the vector potential rather than the scalar potential to extrapolate the initial magnetic field.
   In order to avoid the negative pressure, we modified the energy equation in the similar way 
   as \cite{1999JCoPh.149..270B}.
   This modification violates the energy conservation slightly at the level of discretization errors.
   Using the MUSCL interpolation and the 2nd order Runge-Kutta integration methods, 
   our numerical code achieves the 2nd order accuracy in both space and time.

\section{Quasi Steady Wind Solution}\label{sec:solution}

   \begin{figure}
     \begin{center}
       \includegraphics[scale=1.0]{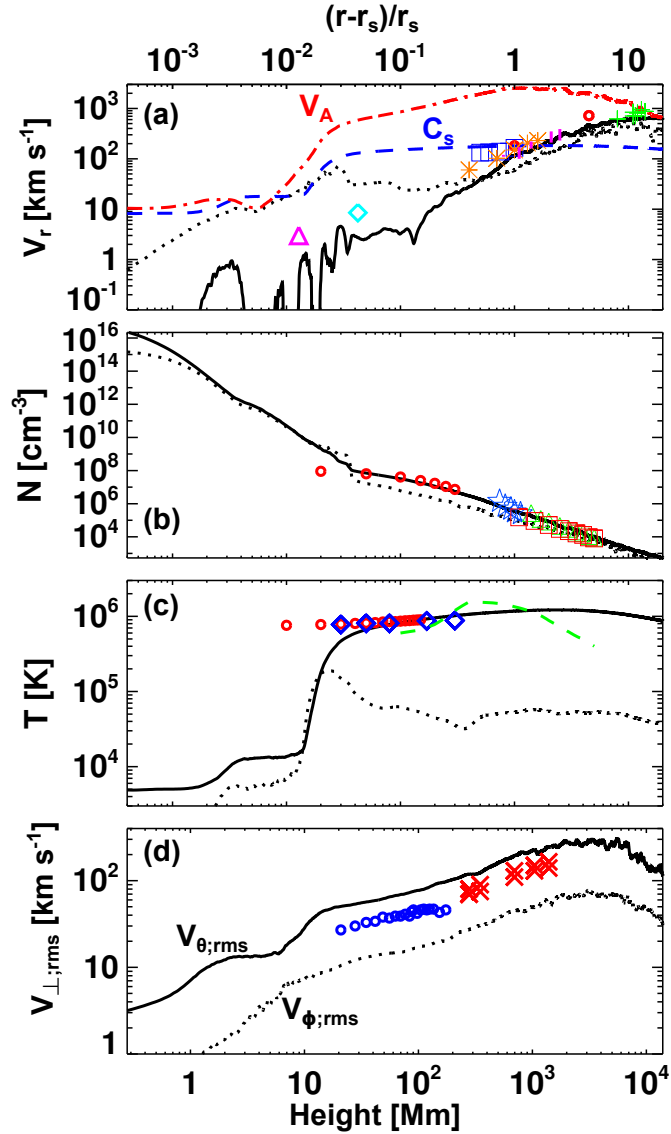}
       \caption{
         Global structure of our numerical simulation after the system has reached the quasi-steady state.
         The black solid line in each panel indicates the mean profile of 
         radial velocity (a), the number density (b), and the temperature (c). The standard deviations
         are superposed as the dotted lines. 
         The blue dashed and red dash-dotted 
         lines in the panel (a) represents sound and Alfv\'{e}n speed, respectively.
         The solid and dotted lines
         in panel (d) represent the root mean square of the velocity in $\theta$ and $\phi$ direction, respectively.
         All the mean profiles are averaged over 100 minutes in temporally and taken along the central 
         axis of the flux tube. The superposed symbols in each panel represent the observed values 
         as follows. 
         (a) The purple 
         triangle, He line onboard SOHO/SUMER \citep{2000A&A...353..749W}.
         The water blue diamond, Ne$^7+$ line onboard SOHO/SUMER \citep{2000A&A...353..749W}.
         The orange asterisks, HI Ly $\alpha$ line onboard SOHO/UVCS \citep{2002ApJ...574..477Z}.
         The blue squares, HI Ly $\alpha$ line onboard SOHO/UVCS \citep{2003ApJ...588..566T}.
         The purple solid bars, HI Ly $\alpha$ line onboard SOHO/UVCS \citep{2000SoPh..197..115A}.
         The red circles, coronagraph on SPARTAN \citep{1995GeoRL..22.1465H}.
         The green pluses, IPS observation with VLBI \citep{1996Natur.379..429G}.
         (b) The red circles, the electron density by SOHO/SUMER \citep{1998ApJ...500.1023W}.
         The blue stars, the electron density by SOHO/UVCS \citep{2003ApJ...588..566T}.
         The green triangles, the electron density by SOHO/LASCO \citep{2003ApJ...588..566T}.
         The orange squares, the electron density by SOHO/LASCO \citep{1997ESASP.404..491L}.
         (c) The red circles, the electron temperature by SOHO/CDS \citep{1999SSRv...87..185F}.
         The blue diamonds, the electron temperature by SOHO/SUMER \citep{1998ApJ...500.1023W}.
         The green dashed line, the electron temperature by SWICS/Ulysses \citep{1997SoPh..171..345K}.
         (d) The blue circles, the non-thermal broadening of Si VIII by SOHO/SUMER \citep{1998ApJ...500.1023W}.
         The red crosses, the non-thermal broadening of Ly $\alpha$ by SOHO/UVCS.
          }
       \label{fig:1dprofile}
     \end{center}
   \end{figure}

   From $t=0$, the foot point of the flux tube is forced to move according to the prescribed velocity 
   perturbation in $\theta$ direction to drive the Alfv\'{e}nic disturbances.
   The system below 20 $r_{\rm s}$ (1.4$\times$10$^4$ Mm) has reached a quasi-steady state after several Alfv\'{e}n crossing times
   have passed. Since we solved time-dependent MHD equations, all the physical variables like
   the density and the temperature could be evolved to adjust the boundary condition.
   We have found that the system has a transonic wind and a thin temperature transition from a cool chromosphere to a hot corona. 
   The reproduction of the transonic wind and the hot corona here is a natural consequence
   of the propagation and dissipation of Alfv\'{e}n wave since we do not assume any prescribed heating or acceleration functions.
   In this section, we will describe the mean structure of our hot coronal wind solution.
   The dynamic features and the detailed heating mechanisms will be described in section \ref{sec:dynamics} and \ref{sec:heating}, respectively.

   The black solid line in figure \ref{fig:1dprofile}a shows the radial velocity as a function of height.
   The radial velocity is taken along the central axis of the flux tube and is temporally 
   averaged over 100 minutes. 
   The dotted line represents the standard deviation at each height.
   The blue dashed and red dash-dotted lines correspond to the sound speed and Alfv\'{e}n speed, respectively.
   The sonic point in our solution is located at $r-r_{\rm s}\approx2$ $r_{\rm s}$ 
   (1.4$\times$10$^3$ Mm) 
   while the Alfv\'{e}n point 
   is located at $r-r_{\rm s} \approx 17$ $r_{\rm s}$ (1.2$\times$10$^4$ Mm). 
   The symbols superposed on the velocity profile represent
   the observational values. The detailed description of the observations are 
   in the caption of figure \ref{fig:1dprofile}.
   The solar wind speed of our model has already reached $\approx$ 600 km s$^{-1}$ at 20 $r_{\rm s}$ (1.4$\times$10$^4$ Mm), 
   which fairly well agree with the observation.

   The temperature and the density profiles are also shown in figure \ref{fig:1dprofile}b and
   figure \ref{fig:1dprofile}c in the similar manner as figure \ref{fig:1dprofile}a.
   The sharp transition layer at $r-r_{\rm s}\approx 0.02$ $r_{\rm s}$ (14 Mm) is located bit higher than 
   the observed transition region height above the coronal hole if we regard the observed
   spicular heights as the transition region height.
   We shall discuss the detailed dynamics of the transition region in section \ref{sec:dynamics}.
   Since we adopt the electron thermal conductivity that is more efficient than that of protons, 
   the temperature in our model can be regarded as the electron temperature.
   
   The solid and the dotted lines in figure \ref{fig:1dprofile}d represent the root mean square
   of $V_\theta$ and $V_\phi$, respectively.
   The superposed symbols correspond to
   the Alfv\'{e}nic fluctuations inferred from the non-thermal line broadening.
   The resultant amplitude of $V_{\theta}$ seems to be about twice larger than the observed amplitude.
   On the other hand,
   the amplitude of $V_\phi$ is significantly smaller than that of $V_\theta$ because
   we impose only Alfv\'{e}nic fluctuations ($V_\theta$) from the photosphere and $V_\phi$ 
   is generated through nonlinear mode coupling from the fluctuation in $\theta$ direction.
   Therefore the expected value of observed velocity ($V_{\rm obs;rms}=[(V_\theta^2+V_\phi^2)/2]^{-1/2}$) could be reduced
   to $2^{-1/2}$ of $V_{\rm \theta;rms}$.
   This effect could be one of the reasons for the discrepancy between the simulated and observed 
   velocity amplitude.

   Figure \ref{initial_field}a shows the magnetic field structure below 10 Mm ($\sim$1.4$\times10^{-2}~r_{\rm s}$) 
   in the quasi-steady state.
   The black/gray solid lines represent open/closed magnetic field lines.
   As is shown in figure \ref{initial_field}a, the photosphere and the chromosphere 
   are filled with strong magnetic field so that non-magnetic atmosphere does not exist in our model.
   Although the flux tube can be deformed according to the resulting gas pressure gradient and the gravity force,
   the change from the initial potential field is not so large in the present simulation.
   The iso plasma beta lines are superposed in figure \ref{initial_field}a as the dashed lines.
   Along the central axis of the flux tube, the plasma beta increases from 0.3 at the bottom due to the rapid expansion of the flux tube.
   The plasma beta ceases to increase at 4.5 Mm with the maximum value of 3 and then decreases with respect to height.
   
\begin{figure}
     \begin{center}
      \includegraphics[scale=1.0]{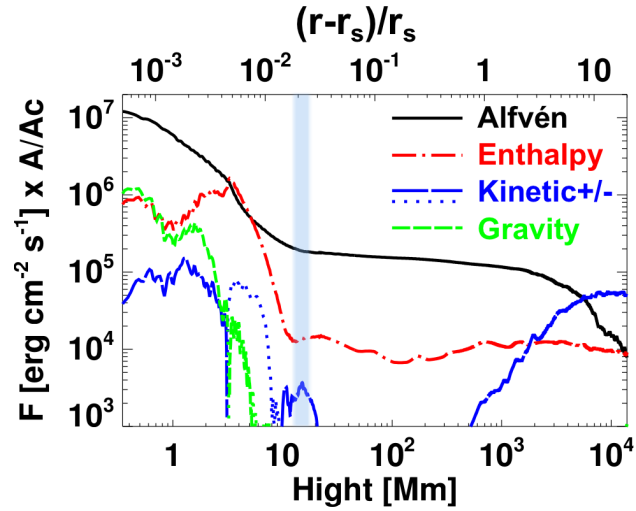}
         \caption{Energy flux in the open field region normalized by the cross section at $r-r_{\rm s}=0.02$ 
                  $r_{\rm s}$ (14 Mm).
                  The black (solid), red (dash-dotted), blue (long-dashed(+) or dotted(-)), 
                  and green (short-dashed) lines represent the Alfv\'{e}nic, enthalpy, and kinetic
                  energy, and gravitational energy flux, respectively.
                  Each energy flux is multiplied by the cross section $A$ of the open flux tube, and
                  normalized by the cross section $A_{\rm c}$ at the coronal bottom $(r-r_{\rm s}=0.02$
                  $r_{\rm s}$ or 14 Mm).
                  The blue shaded area corresponds to the transition region.
                  }
         \label{energy_flux}
     \end{center}
   \end{figure}
\begin{figure}
     \begin{center}
      \includegraphics[scale=1.0]{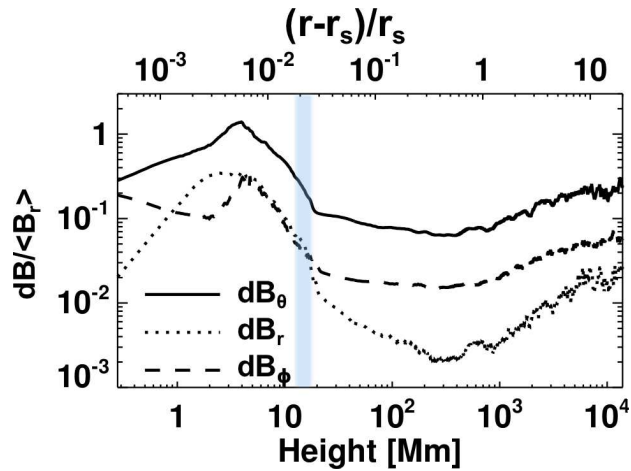}
         \caption{Nonlinearity of magnetic field disturbance along the central axis of the flux tube
                  with respect to height. The solid, dotted, and dashed lines correspond to
                  $\sqrt{\langle B_\theta^2 \rangle}$,$\sqrt{\langle (B_r-\langle B_r \rangle)^2 \rangle}$, 
                  and $\sqrt{\langle B_\phi^2 \rangle}$ normalized by $\langle B_r \rangle$, respectively.
                  The blue shaded area corresponds to the transition region.
                  }
         \label{fig:nonlinearity}
     \end{center}
   \end{figure}

   Figure \ref{energy_flux} shows the energy flux distribution along the flux tube.
   The mechanical energy flux that transports the energy outwardly is composed of the following components,
   \begin{eqnarray}
     F_{\rm alf} &\equiv& \left( {1\over 2} \rho V_{\perp}^2 + B_\perp^2 \right) V_{r} 
     - \bmath{V_{\perp}} \cdot \bmath{B_{\perp}} B_{r}, \\
     F_{\rm ent} &\equiv& {\gamma \over \gamma -1} P_{\rm g} V_{r}, \\
     F_{\rm knt} &\equiv& {1\over 2} \rho V_{r}^3 , \\
     F_{\rm grv} &\equiv& -{\rho V_r GM \over r},
   \end{eqnarray}
   where $F_{\rm alf}, F_{\rm ent}, F_{\rm knt}$, and $F_{\rm grv}$ are defined to represent 
   the Alfv\'{e}nic, enthalpy, kinetic, and gravitational energy flux, respectively.
   The energy flux in figure \ref{energy_flux} is multiplied by the cross section $A$ of the open flux tube.
   After that, we normalized the energy flux by the cross section $A_{\rm c}$ at the coronal bottom 
   $(r-r_{\rm s}=0.02$ $r_{\rm s}$ or 14 Mm).
   The Alfv\'{e}n wave flux is well above $10^5$ erg cm$^{-2}$ s$^{-1}$, a reasonable value to maintain the hot corona
   \citep{1977ARA&A..15..363W}.
   It is found that the Alfv\'{e}n wave flux dominates the other two energy flux except in two different regions.
   The enthalpy energy flux becomes comparable to the Alfv\'{e}n wave energy flux at around $r-r_{\rm s}=0.004$ 
   $r_{\rm s}$ (3 Mm), 
   the region where the flux tube is just opened up.
   The Alfv\'{e}nic fluctuation at the photosphere is converted to the enthalpy flux 
   through the linear/nonlinear mode conversion \citep{2012ApJ...749....8M}.
   Since the wave amplitude is large below the transition region as shown in figure \ref{fig:nonlinearity}, 
   the efficient nonlinear mode conversion can occur there.
   The solid, dotted, and dashed lines in figure \ref{fig:nonlinearity} correspond to
   $\sqrt{\langle B_\theta^2 \rangle}$,$\sqrt{\langle (B_r-\langle B_r \rangle)^2 \rangle}$, 
   and $\sqrt{\langle B_\phi^2 \rangle}$ along the central axis of the flux tube normalized by $\langle B_r \rangle$ 
   , respectively.   
   Although our numerical results would idealy be axisymmetric, 
   the perturbations of $B_\phi$ appear even along the central axis.
   This is because the initial magnetic field structure is not perfectly axisymmetric numerically and
   causes kink motions ($B_\phi$) of the flux tube.
   Above $r-r_{\rm s}\sim7$ $r_{\rm s}$ (5$\times$10$^3$ Mm), the kinetic energy flux exceeds the Alfv\'{e}n wave flux.
   The Alfv\'{e}n wave pressure pushes the ambient plasma to accelerate the solar wind.

   \begin{figure}
     \begin{center}
       \includegraphics[scale=1.0]{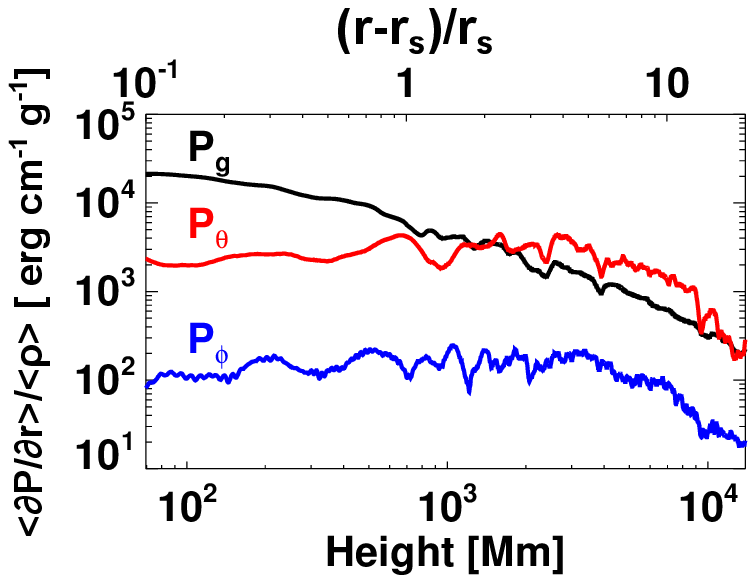}
       \caption{
               Pressure gradient force per unit mass as a function of height.
               The black, red, and blue lines represent 
               the contribution from gas pressure ($P_g$), and wave pressure ($P_{\theta,\phi}=B_{\theta,\phi}^2/2$),
               respectively.
               The profiles are averaged over 100 minutes temporally and over $r=r\pm0.1r$ spatially.
               }
       \label{fig:pr_sample}
     \end{center}
   \end{figure}

   The solar wind in our simulation is accelerated by the wave pressure in addition to the gas pressure.
   Figure \ref{fig:pr_sample} reveals the pressure gradient force with respect to height.
   The contribution from the gas pressure ($P_{g}$), and the wave pressure ($P_{\theta,\phi}=B_{\theta,\phi}^2/2$) 
   is plotted as the black, red, and blue solid lines.
   Above the sonic point ($r-r_{\rm s}>2$ $r_{\rm s}$ or 1.4$\times$10$^3$ Mm),
   the contribution from the wave pressure ($P_{\theta}$) exceeds 
   that from the gas pressure ($P_g$). 
   Then at around the Alfv\'{e}n point ($r - r_{\rm s}\approx 17$ $r_{\rm s}$ or 1.2$\times$10$^{4}$ Mm), 
   the contributions from $P_g$ and $P_{\theta}$ become comparable again.
   The $\phi$ component of the wave pressure ($P_{\phi}$) is always less effective
   than the other two components.
   

\section{Dynamic features}\label{sec:dynamics}

   In this section, we shall describe the dynamic features in our simulation.
   Before going into the detailed discussion, the terminology of the wave mode 
   and the linear property of our model should be clarified.
   Since $B_{\theta}$ averaged over time is nearly zero everywhere,
   $(V_{\theta}, B_{\theta})$ and 
   the other variables ($\rho, P_g, V_{r,\phi}, B_{r,\phi}$) are almost decoupled within linear theory.
   We shall call any perturbations of $V_{\theta},B_{\theta}$ as the Alfv\'{e}nic component.
   Perturbation of $\rho, P_g, V_{r,\phi}, B_{r,\phi}$ are referred to as the compressible component.
   The compressible component could be decomposed into slow and fast mode 
   and they can be coupled each other even in linear theory 
   in a non-uniform medium.
   \citep[e.g.][]{2003ApJ...599..626B}.
   In our simulation, the coupling between the Alfv\'{e}nic and the compressible component should be nonlinear one 
   that is due to the wave pressure, or ponderomotive force \citep[e.g.][]{2013SoPh..288..205T}.

   \subsection{Dynamics below the transition region}

   \begin{figure*}
     \begin{center}
      \includegraphics[scale=1.0]{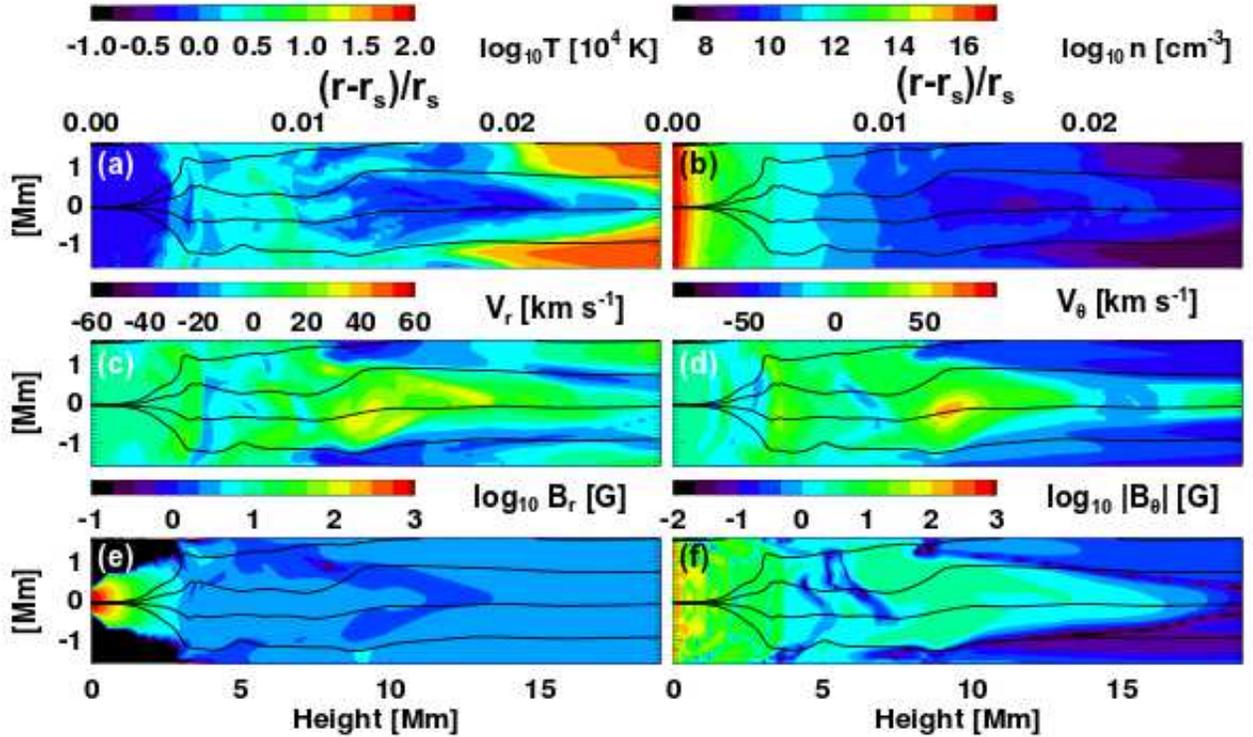}
         \caption{
           Snapshot images of our simulation after the system has reached 
           the quasi steady state. The background colors indicate the temperature (a),
           the number density (b), the radial velocity (c), $V_\theta$ (d),
           the radial magnetic strength (e), and the absolute value of the 
           $B_\theta$ (f). The black lines in each panel represent
           the magnetic field lines. Only the open field lines are shown here.
         }
         \label{fig:2dsnapshot}
     \end{center}
   \end{figure*}

   Figure \ref{fig:2dsnapshot} (video available online, movie 1)
   shows the snapshot images of 
   the temperature (a), the number density (b), the radial velocity (c),
   $V_\theta$ (d), the radial magnetic field strength (e),
   and $|B_\theta|$ (f).
   The black lines in each panel represent the magnetic field lines projected onto $r-\phi$ plane.
   Note that the aspect ratio is not correctly displayed in this figure.
   
   Although the prescribed Alfv\'{e}nic perturbation is uniform in $\phi$ direction, 
   the structures of the higher atmosphere are significantly inhomogeneous in $\phi$ direction.
   This is mainly because of the inhomogeneity in the magnetic field.
   The Alfv\'{e}n speed along the central axis of the flux tube is generally larger than its surroundings.
   Even when the initial Alfv\'{e}n wave front is straight in $\phi$ direction, the central part of the 
   wave front tends to propagate faster than the off axis part.
   This leads to the curved wedge or convex shape of the Alfv\'{e}n wave front, 
   as was demonstrated by \cite{1997ApJ...488..854C}.
   This inhomogeneous Alfv\'{e}nic disturbance creates the radial velocity 
   through the nonlinear mode conversion \cite[e.g.][]{1982SoPh...75...35H}, resulting in the inhomogeneity in the 
   density and the temperature.

   \subsection{Time distance diagram from Photosphere to Low Corona}

\begin{figure*}
     \begin{center}
       \includegraphics[scale=1.0]{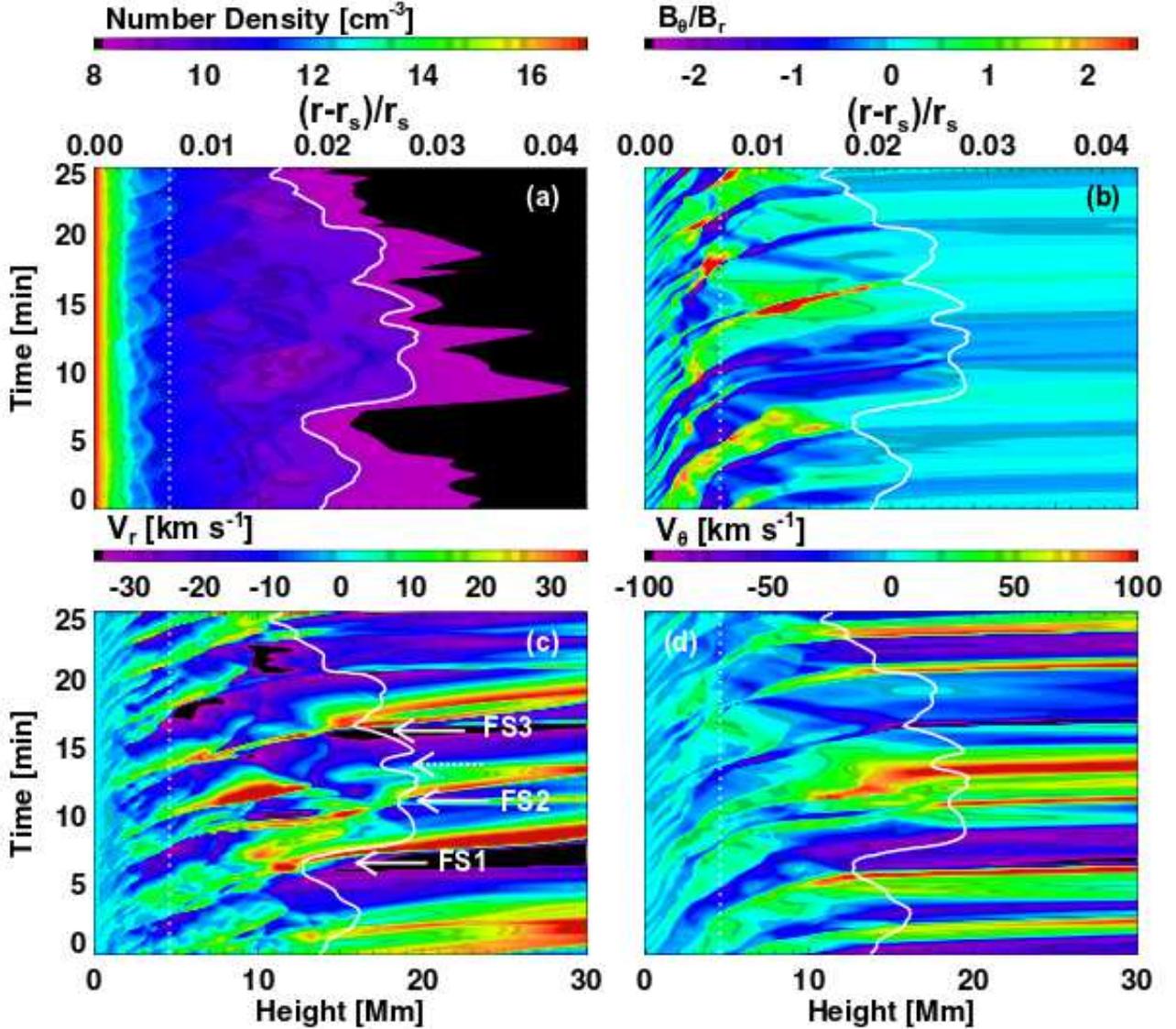}
       \caption{Time distance diagram along the central axis of the flux tube below 30 Mm. 
         The background colors in each panel indicate the number density (a), 
         the Alfv\'{e}n wave nonlinearity ($B_\theta/B_r$) (b), the radial velocity (c),
         and $V_\theta$ (d). The white solid line in each panel represents
         the transition layer where the temperature is 10$^5$K.
         The white dotted line in each panel indicates the height just above the magnetic canopy
         averaged over 100 minutes.
           FSs with solid arrows in the panel (c) indicate the collision events 
           between the fast shock and the transition region.
           FS with dotted arrow in the panel (c) corresponds to the swaying motion of the transition region.
       }
       \label{fig:td_diagram_near}
     \end{center}
   \end{figure*}

   In order to understand the wave propagation and its interaction with the transition region,
   time distance diagrams are convenient.
   Here we focus on dynamic features below 30Mm (4.3$\times$10$^{-2}$ $r_{\rm s}$) from the photosphere.
   Figure \ref{fig:td_diagram_near} represents time distance diagrams of the number density (a),
   the Alfv\'{e}n wave nonlinearity or $B_{\theta}/B_{r}$ (b), the radial velocity (c), and 
   $V_\theta$ (d). We took the values along the central axis of the flux tube as the horizontal axis 
   and stack the time variance of them onto the vertical axis.
   The white solid line in each panel corresponds to the transition region height where the temperature is 10$^5$K.
   The white dotted line in each panel indicates the height just above the magnetic canopy averaged over 100 minutes.
   We can observe a lot of oblique signatures of wave propagation whose inclinations correspond to phase speed.
   The phase speed increases both in compressible (panels a \& c) and Alfv\'{e}nic (panels b \& d) fluctuations at the transition region.
   However, only the phase speed in Alfv\'{e}nic fluctuations increases just above the magnetic canopy.
   Since the super radial decrease in the magnetic field strength is suppressed just above the canopy by the surrounding flux tubes, 
   Alfv\'{e}n speed increases exponentially according to the decrease in the mass density.

   The transition region height exhibits up-and-down motions.
   Many authors suggest the similarity between these motions and the solar spicules
   \citep{1982SoPh...75...35H,1999ApJ...514..493K}.
   In figure \ref{fig:td_diagram_near}c, we have at least 4 ascending motions of the transition region 
   indicated by white arrows.
   The ascending events indicated by the solid arrows are associated with upward velocity.
   This upward velocity originates from fast shocks that could be driven by the nonlinear steepening of 
   Alfv\'{e}n waves.
   In low beta plasma such as the chromosphere (around the height of 10 Mm or 1.4$\times$10$^{-2}$ $r_{\rm s}$) 
   in our simulation, 
   the nonlinear Alfv\'{e}n waves produce switch-on shocks as well as slow mode waves. 
   This switch-on shocks lift up the transition region.
   The daughter slow mode waves can also steepen into shocks due to the gravitational stratification
   although we can not see the ascending events due to the slow shocks in this time span.

   The ascending event indicated by the dotted arrows are not associated with upward velocity.   
   This is just an apparent motion due to the swaying motion of the transition region.
   We took the time distance diagram along the axis of the flux tube.
   The transition region is lifted by the fast shock (FS2 in figure \ref{fig:td_diagram_near}c) 
   that makes the transition region wedged shape
   which is similar to the shape shown in figure \ref{fig:2dsnapshot}a.
   The wedged shape transition region reveals swaying motion in $\phi$ direction.
   If the swaying motion is so large we can see the apparent up-and-down motion of the transition region in 
   the time distance diagram.

   \subsection{Time distance diagram from Corona to Solar Wind}

\begin{figure}
      \begin{center}
      \includegraphics[scale=1.0]{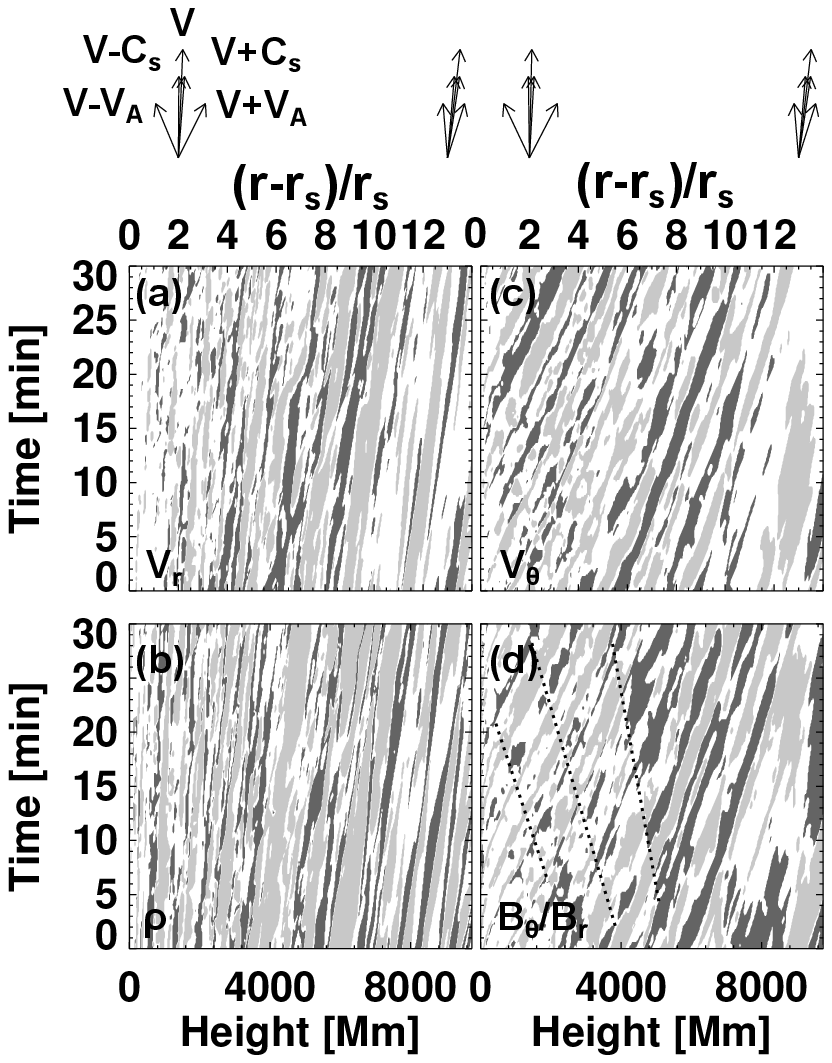}
         \caption{Time distance diagram along the central axis of the flux tube below 15 $r_{\rm s}$ 
                 (1.0 $\times$ 10$^4$ Mm).
                 (a) The variations of the radial velocity, $V_{r} - \langle V_{r} \rangle, $ 
                 that exceed the threshold of +/- 60 km s$^{-1}$ are shown in dark/light contours.
                 (b) The relative variations of the density, $(\rho - \langle \rho \rangle)/\langle \rho \rangle$.
                 The thresholds are +/- 0.25.
                 (c) The variations of $V_{\theta}$.
                 The thresholds are +/- 180 km s$^{-1}$.
                 (d) The relative variations of $B_{\theta}/\langle B_{r} \rangle$.
                 The thresholds are +/- 0.1.
                 The dotted lines corresponds to the signature of reflection waves.
                 The mean phase velocity, $V_{r},V_{r}\pm C_{\rm s},V_{r}\pm V_{A}$ are 
                 plotted as arrows on the top of the panel as references.
                  }
         \label{fig:td_diagram_far}
      \end{center}
   \end{figure}

   Here we focus on wave phenomena up to 15 $r_{\rm s}$ (1.0 $\times$ 10$^4$ Mm).
   Figure \ref{fig:td_diagram_far} shows the time distance diagrams along the central axis of the flux tube below 15 
   $r_{\rm s}$ (1.0 $\times$ 10$^4$ Mm).
   The radial velocity, $V_{r} - \langle V_{r} \rangle$ (a), the density, $( \rho - \langle \rho \rangle ) / \langle \rho \rangle$ (b), 
   $B_{\theta}/\langle B_{r} \rangle$ (c), and $V_{\theta}$ (d) are 
   plotted as contours when the values exceed a certain threshold.
   The thresholds are $\pm$ 60 km s$^{-1}$ (a), $\pm$ 0.25 (b), $\pm$ 180 km s$^{-1}$(c), and $\pm$ 0.1 (d).
   The average operator $\langle \rangle$ here means temporal average over 100 minutes.
   The mean phase velocity, $V_{r},V_{r}\pm C_{\rm s},V_{r}\pm V_{A}$ are 
   plotted as arrows on the top of the panel as references.

   The propagating signature in the radial velocity and the density fluctuations in figure \ref{fig:td_diagram_far} 
   corresponds to compressible waves.
   From the slope of the signature, these compressible waves are considered to be slow mode waves.
   There are several origin of the slow mode wave in the corona.
   In the chromosphere, the nonlinear steepening of Alfv\'{e}n waves produces the switch-on (fast) shocks.
   When the switch-on shocks enter the corona they creat slow mode shocks as well as fast shocks.
   The nonlinear mode conversion or the parametric decay of the Alfv\'{e}n waves in the corona can also be 
   the origin of the slow mode waves since the nonlinearity of the Alfv\'{e}n waves is not so small 
   (0.1-0.2) in the corona (Fig. \ref{fig:nonlinearity}).
   The density fluctuation reaches $\approx$ 0.8 on average at $r=10$ $r_{\rm s}$ (7.0 $\times$ 10$^3$ Mm) 
   and sometimes exceeds unity.
   The fluctuation in Alfv\'{e}n speed due to the density fluctuation could be the reflection source of
   the Alfv\'{e}n waves.

   The figure \ref{fig:td_diagram_far}c and \ref{fig:td_diagram_far}d represents the propagation of the Alfv\'{e}n waves.
   Besides the out going Alfv\'{e}n waves, the signatures of reflected waves 
   can be seen as the dotted lines
   in the figure \ref{fig:td_diagram_far}d.
   This is due to the density fluctuation by the slow mode waves and 
   to the gradual change in the background Alfv\'{e}n speed \citep{2005ApJ...632L..49S,2006JGRA..11106101S}.
   The reflection of Alfv\'{e}n waves are important in the context of the solar wind turbulence.
   The wave-wave interaction between out and in going Alfv\'{e}n waves are considered to trigger the nonlinear 
   cascade of MHD turbulence \citep{1999ApJ...523L..93M}.
   Although MS12 suggested that the anisotropic cascade can be seen in their simulation, 
   improving the radial grid size removed the signature of the cascade.
   We probably need full 3 dimensional simulations to describe the nonlinear wave-wave interaction
   \citep{2011ApJ...736....3V}.

\section{Data Analysis} \label{sec:data_analysis}

  Before investigating heating mechanisms that actually work in our model, 
  we should describe how to estimate the heating rate.
  Since we do not have any explicit viscosity and resistivity in our basic equations, 
  the estimation of the heating rate is not so straightforward task.
  Inclusion of explicit dissipation terms will help us to estimate the heating rate
  although it will smooth out large scale structures as well as small scale structures.
  Hyper diffusion method will keep large scale structures sharp 
  while appropriate dissipation is introduced in small scale \citep{2008ApJ...677.1348R,2011ApJ...736....3V}.
  However, the appropriate diffusion coefficients may vary as a function of grid size and typical velocity such as 
  Alfv\'{e}n and sound speed, 
  it is not straightforward to choose the functional form of the diffusion coefficients.
  Therefore instead of using hyper diffusion, we simply choose not to use explicit dissipation terms as a first step, 
  although the estimation of heating rate will be more complicated and less rigorous.

  In this section, we briefly explain the idea to estimate the total heating rate.
  Basically there are two kinds of dissipation in numerical simulations without explicit physical dissipation terms.
  The first one is physical dissipation associated with shocks.
  This dissipation is introduced when Riemann solver is used for flux estimation while the artificial viscosity 
  may be used for other methods.
  The second one is numerical dissipation associated with truncation errors.
  Advection of shear flows or magnetic shear could cause the numerical dissipation.
  This type of dissipation is not always physical dissipation especially when the
  numerical resolution is very poor.
  This type of numerical dissipation could be regarded as the
  physical one if there is the physical reason for small scale cascading such as turbulence.

  For an illustrative purpose, we here explain our method to measure the heating rate in the 1.5 MHD 
  in Cartesian coordinate
  system without thermal conduction and radiative cooling as a simplest example.
  The 1.5 dimensional system here means the system that has spatial variation only in $x$ direction but has vector component in $y$ direction in addition to $x$ direction.
  Please see appendix \ref{app:method} for the actual method we are performing in our spherical 2.5D MHD system.
  
  Since we use the finite volume method, the internal energy $u$ will be 
  calculated after all the conservative variables are updated.
  Then the discretized form of the time derivative of $u$ can be written down as follows.
  \begin{eqnarray}
    {u^{n+1}_i - u^n_i \over \Delta t} = {E^{n+1}_i-E^n_i \over \Delta t} 
    - {1\over \Delta t}\left\{{(M^{n+1}_i)^2 \over2 \rho^{n+1}_i} - {(M^n_i)^2 \over2 \rho^n_i}\right\} 
    - {1\over \Delta t}\left\{{(B^{n+1}_i)^2 \over 2} -{(B^{n}_i)^2 \over 2} \right\} \label{eq:deltan},
  \end{eqnarray}
  where $i$ and $n$ indicate the spatial and the temporal index, respectively.
  The momentum in Cartesian coordinate system, $M_{x,y}$, corresponds to ${\cal M}_{r,\phi}$ in eq (\ref{eq:cnsvar}) and
  defined to be $\rho V_{x,y}$.
  If we transform the time differences of the conservative variables on the right hand side of eq (\ref{eq:deltan}) to the spatial differences of the corresponding flux, 
  \begin{eqnarray}
    {u^{n+1}_i - u^n_i \over \Delta t} &=& Q_{n;i} + Q_{a;i}, \label{eq:decomp}
  \end{eqnarray}
  where
  \begin{eqnarray}
    Q_{n;i} &=& -\Delta_x \left\{(u+P_g)V_x \right\} +{M^{n+1/2}_{x;i} \over \rho^{n+1}_i} \Delta_x P_g
  \end{eqnarray}
  and
  \begin{eqnarray}
    Q_{a;i}&=&{M^{n+1/2}_{x;i} \over \rho^{n+1}_i} \Delta _x \left(\rho V_x^2\right)+{M^{n+1/2}_{y;i} \over \rho^{n+1}_i} \Delta _x \left(\rho V_xV_y\right) \nonumber \\
    &-& {(M^n_i)^2 \over 2 \rho^n_i \rho^{n+1}_i} \Delta_x \left( \rho V_x \right)- \Delta_x \left( {1\over 2} \rho V^2 V_x \right) \nonumber \\
    &+& {M^{n+1/2}_{x;i} \over \rho^{n+1}_i} \Delta_x {B_y^2 \over 2} - {M^{n+1/2}_{y;i} \over \rho^{n+1}_i} \Delta_x {B_x B_y} \nonumber \\ 
    &-& B_{y;i}^{n+1/2} \Delta_x \left( B_x V_y- B_y V_x \right) - \Delta_x \left( B_y^2V_x -V_yB_yB_x\right).
  \end{eqnarray}
  The spatial difference are defined as $\Delta_x f \equiv (f^*_{i+1/2}-f^*_{i-1/2})/\Delta x_i$ and the asterisk on $f$ indicate the variables at the cell surface estimated by the approximate Riemann solver.
  $f^{n+1/2}_i$ is defined to be $(f^{n+1}_i+f^{n}_i)/2$ for arbitral variable $f$.
  Note that $B_x$ is constant in 1D Cartesian case and can be removed from inside the difference.
  The first term in eq (\ref{eq:decomp}), $Q_{n}$, is generally non zero term and roughly corresponds to the discretized form of the adiabatic expansion (-$\gamma u \partial u / \partial x$) plus
  the advection ($-V_x\partial u / \partial x$). This term, however, is slightly different from the adiabatic expansion and advection term since the numerical flux used in the spatial difference 
  may implicitly have dissipative component originated from the (approximate) Riemann solver. 
  The second term in eq (\ref{eq:decomp}), $Q_{a}$, should analytically equal to zero when we replace the spatial difference with the spatial derivative.
  However, discretizing operation leads $Q_{a}$ to have some residual values that we consider contribute partly to the numerical dissipation.
  $Q_{n}$ roughly corresponds to the sum of adiabatic heating and heating 
  at hydrodynamical shocks, while $Q_{a}$ consists of the rest of all the 
  entropy generation, the sum of numerical viscous dissipation by velocity 
  shear and numerical resistive dissipation of magnetic field. 

  Test simulations in appendix \ref{app:method} suggests that $Q_{a}$ always gives good estimation for Alfv\'{e}n waves
  while $Q_{a}$ is good indicator for fast waves only in low beta plasma.
  For slow waves, $Q_{a}$ always underestimates numerical dissipation significantly and 
  the dissipation mainly originates from $Q_{n}$.

  Generally $\langle Q_{n} \rangle$ can be divided into two components.
  The first one is organized only by bulk variables such as $\langle u \rangle$, $\langle P \rangle$, $\langle V_{\rm x} \rangle$.
  This component could be considered as the adiabatic loss due to the solar wind expansion.
  The second one is organized by the cross correlation between fluctuations in $u,P,V_{\rm x}$.
  This component possibly originates from the heating due to acoustic and shock waves.
  Accordingly, the solar wind loss component ($Q_{\rm WL}$) and the acoustic and shock wave component ($Q_{\rm AS}$) 
  can be described as follows.

  \begin{eqnarray}
    Q_{\rm WL} &\equiv& - {\partial \over \partial x} \left\{ (\langle u \rangle + \langle P_g \rangle ) \langle V_{\rm x} \rangle \right) 
    + \langle V_x \rangle \langle {\partial P_g \over \partial x} \rangle, \\
    Q_{\rm AS} &\equiv& \langle Q_{n} \rangle - Q_{\rm WL}.
  \end{eqnarray}
  Note that $Q_{\rm WL}$ and $Q_{\rm AS}$ are the temporally averaged value while $Q_{n}$ and $ Q_{a}$ are the functions of time.

\section{Heating Rates}\label{sec:heating}

\begin{figure}
     \begin{center}
      \includegraphics[scale=1.0]{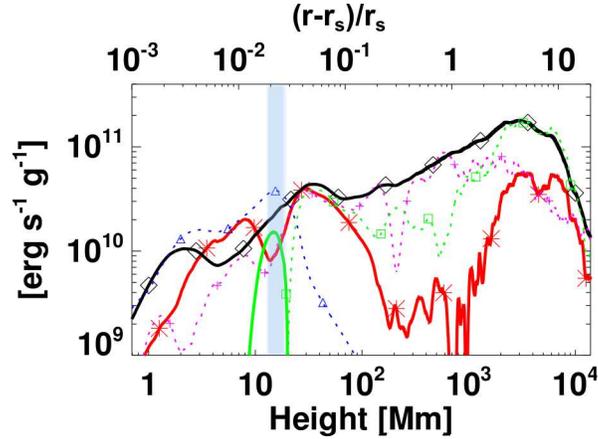}
         \caption{
                 Energy balance after the system has reached the quasi-steady state.
                 The black (with diamonds), green (with squares), blue (with triangles)
                 , red (with asterisks), and purple (with pluses) lines correspond to
                 the $Q_{a}$, the thermal conduction,
                 the radiative loss, $Q_{\rm AS}$ and the adiabatic loss due to the solar wind ($Q_{\rm WL}$), respectively.
                 All the heating/cooling rates are normalized by $\langle \rho \rangle$ in this plot.
                 The solid lines indicate the positive value (heating) while the 
                 dotted lines indicate the negative value (cooling).
                 The blue shaded area corresponds to the transition region.
                 }
         \label{heating_per_unitmass}
     \end{center}
   \end{figure}

   In section \ref{sec:data_analysis}, we describe the idea to precisely measure
   the heating rate that actually occurred in our simulation.
   Figure \ref{heating_per_unitmass} represents the energy balance with respect to height.
   All the heating rates below are averaged over 100 minutes temporally and over all $\phi$ domain spatially.
   Moreover, the heating rates are spatially averaged over $0.8(r-r_{\rm s})<r-r_{\rm s}<1.2(r-r_{\rm s})$.
   The black solid line with diamonds corresponds to $Q_{a}$ averaged over time and space.
   The green solid/dashed line with squares indicates the conduction heating/loss while the blue dashed line
   with triangles
   shows the radiation loss. $Q_{\rm AS}$ is represented by the red solid line with asterisks while 
   the solar wind loss or $-Q_{\rm WL}$ is denoted by the purple dashed line with pluses.
   Below the transition region ($r-r_{\rm s}<0.02$ $r_{\rm s}$ or 14 Mm), 
   the radiative loss is balanced by the sum of $Q_{\rm AS}$ and $Q_{a}$, or mechanical heating rate.
   Around the transition region ($r-r_{\rm s}\approx0.02$ $r_{\rm s}$ or 14 Mm), 
   the contribution of the thermal conductive heating
   becomes significant as well as the radiation and the mechanical heating.
   The sudden increase in temperature profile causes the strong increase of the thermal conductive heating here.
   At the coronal bottom $r-r_{\rm s}\approx0.05$ $r_{\rm s}$ (35 Mm), 
   the thermal conductive loss and the solar wind loss are balanced by the mechanical heating rate
   and the radiative cooling becomes negligible.
   In the solar wind acceleration region ($r-r_{\rm s}>r_{\rm s}$ or 700 Mm), 
   the thermal conduction loss is balanced by $Q_{a}$.

   \subsection{Heating below the transition region}

\begin{figure}
     \begin{center}
      \includegraphics[scale=1.0]{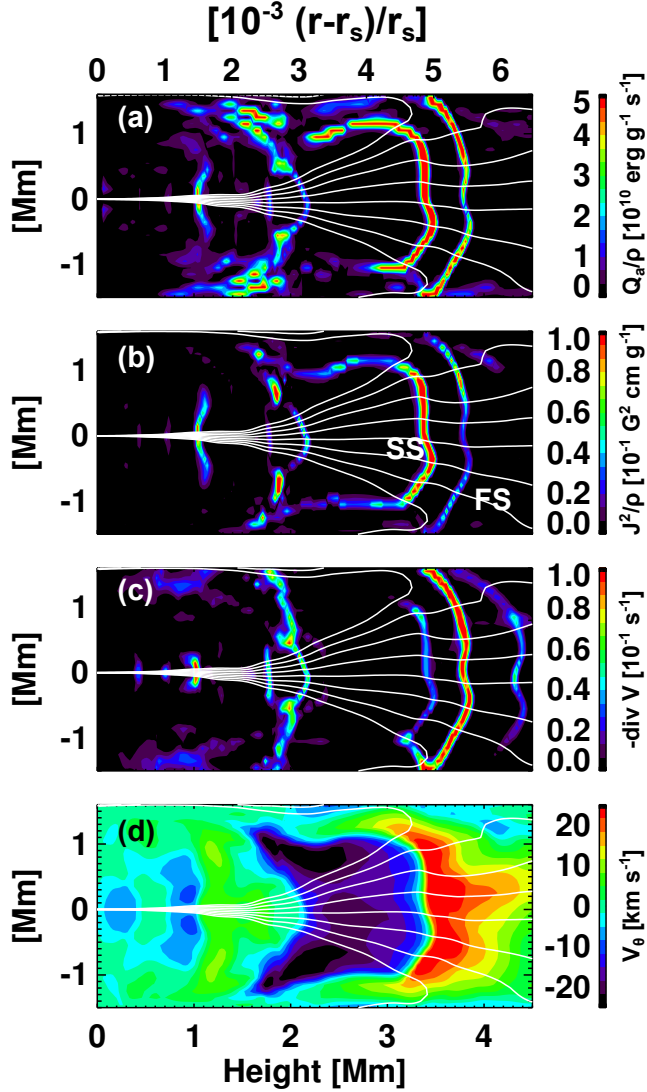}
         \caption{Heating events around the chromosphere. 
                 The color contours represent $Q_{a}/\langle \rho \rangle$ (a),
                 the current density square per unit mass (b), 
                 the velocity convergence ($\nabla\cdot v<0$) (c),
                 and $V_\theta$ (d).
                 The white lines in each panel indicate the magnetic field lines.
                 }
         \label{fig:heating_below_4Mm}
     \end{center}
   \end{figure}

   The mean profiles of the heating rate clearly show that the heating events below the transition region 
   are dominated by $Q_{a}$ and $Q_{\rm AS}$ (fig \ref{heating_per_unitmass}).
   Combining the mean heating profiles and the snapshot images of the total heating rate,
   we found that the dominant heating mechanism below the transition region was shock heating.
   Figure \ref{fig:heating_below_4Mm} (video available online, movie2) 
   shows the snapshot images of $Q_{a}/\langle \rho \rangle$ (a), 
   the squared current density over the mass density (b), the velocity convergence ($\nabla\cdot v<0$) (c), and
   $V_\theta$ (d).
   The spatial distribution and the time evolution of the heating rate 
   tell us what is the actual heating mechanisms in our numerical simulation.
   There is a shock (FS in figure \ref{fig:heating_below_4Mm}b) propagating toward right hand side. 
   At the down stream of the shock, the magnetic pressure ($B_\theta^2/2$) increases.
   From this property, we identified FS as a fast shock.
   The location of FS corresponds to the region where the heating rate ($Q_a$) is significantly high.
   The locations with converging motion ($\nabla \cdot v<0$) tend to have high $Q_a$ in the chromosphere, 
   which means the chromospheric shocks are well captured by $Q_a$.
   We also have high $Q_a$ region without conversing motion probably because of the low numerical resolution.
   At most 2/3 of $Q_a$ comes from the shock region in the chromosphere (fig \ref{fig:cmp_icmp_compare}a), 
   although the ratio will be smaller in higher numerical resolution.
   The contribution of the slow shocks could be important as well.
   There are also a lot of switch off slow shocks in our simulation, for example SS in figure \ref{fig:heating_below_4Mm}b.
   Instead of the swith off shocks, the switch on fast shocks will appear 
   at the higher height where the background plasma beta is low.
   As was discussed by \cite{1982ApJ...254..806H}, the fast switch on shocks are
   one of the main mechanisms in our simulation in the higer chromosphere.
   We think both the fast and slow shocks are important heating mechanisms in our simulation, 
   although we have not yet elucidated which is more important statistically.
   
   \subsection{Heating just above the transition region}

\begin{figure*}
     \begin{center}
      \includegraphics[scale=1.0]{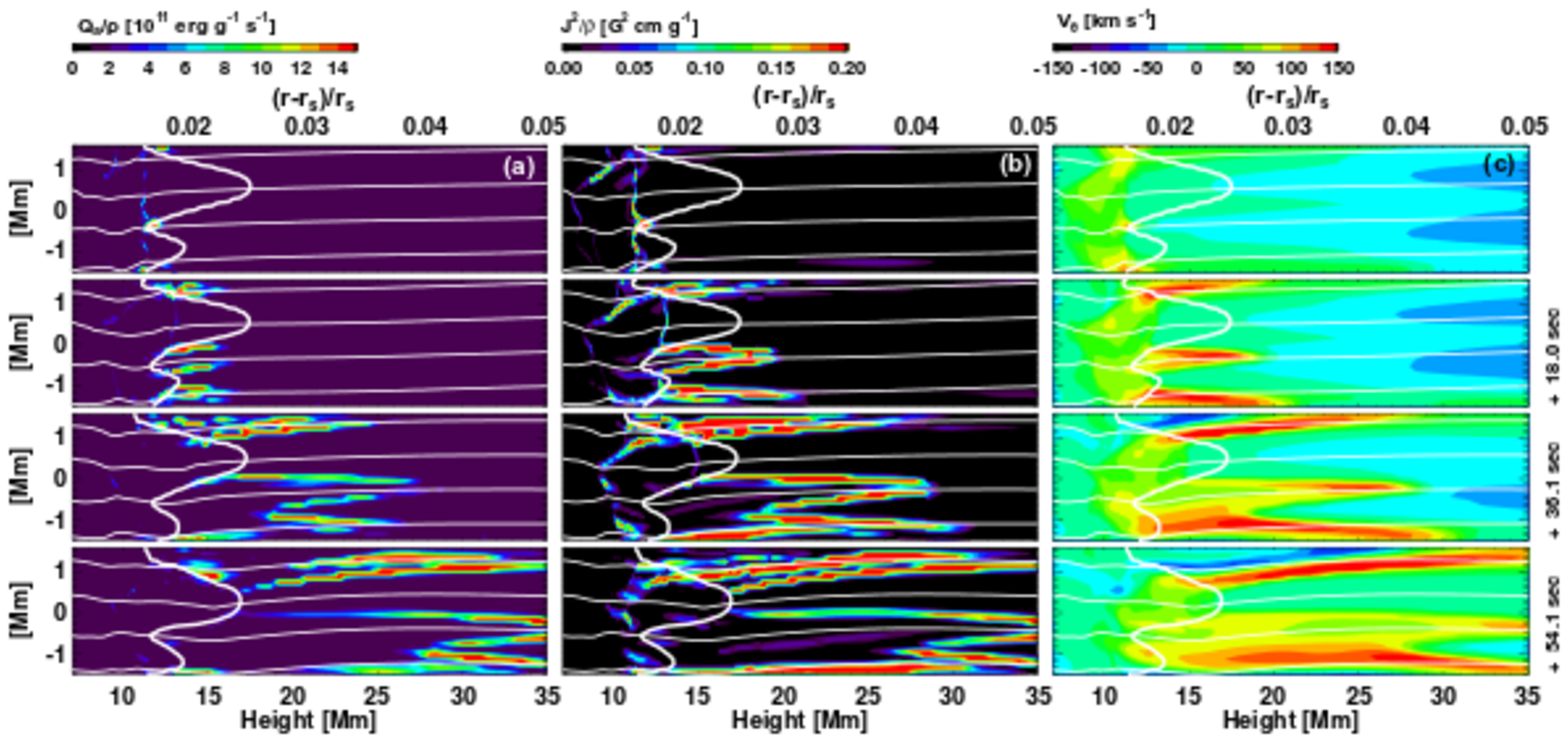}
         \caption{Time series of a collision event between an Alfv\'{e}n wave and the transition region.
                 The left (a), middle (b), and right (c) column represent $Q_{a}/\langle \rho \rangle$
                 , the current density square per unit mass, and $V_\theta$, respectively. 
                 The thin white lines indicate the magnetic field lines. 
                 The thick white line in each panel shows the transition layer where the temperature is 10$^5$ K.
                 }
         \label{fig:heating_above_tr}
     \end{center}
   \end{figure*}

   The heating mechanism just above the transition region is different from that below the transition region.
   Figure \ref{fig:heating_above_tr} shows the temporal evolution of a typical heating event just above
   the transition region (video available online, movie 3). 
   From the left to the right column, $Q_{a}/ \langle \rho \rangle$, 
   the squared current density per unit mass, and $V_\theta$ are shown.
   Time goes on from the top to the bottom with 18 second temporal intervals.
   The transition region is indicated by the thick white lines and is significantly deformed due to collisions with shocks.

   In the first row of figure \ref{fig:heating_above_tr}, an Alfv\'{e}n wave front comes from the left hand side.
   The injected Alfv\'{e}n wave has already been nonlinearly steepened into the fast shock to have small length scale in
   radial direction before they collide with the transition region.
   The steepening in radial direction creates the current density in $\phi$ direction along the wave front as was seen
   in the first row of figure \ref{fig:heating_above_tr}.
   Between $t=0$ and $t=+18$, the fast shock collides with the transition region.
   The shock front refracts significantly 
   because of the huge difference in phase speed across the transition region.
   When the chromospheric fast shock hits the transition region, the chromospheric fast shock
   would split into an out-going fast/slow shock, a contact discontinuity ( can be seen as spicules )
   and an in-going fast rarefaction wave.
   Since the Mach numbers of the fast shock are so small that we could not see the shocks clearly.
   In addition, we have an out-going Alfv\'{e}n wave between the fast and slow shock.
   The resultant Alfv\'{e}n wave has an elongated wedge shape structure 
   shown in figure \ref{fig:heating_above_tr}c.
   We do not detect velocity convergence ($\nabla\cdot v<0$) along the wedge shape structure, 
   which means this structure is not the shock but some kind of Alfv\'{e}n waves.
   The Alfv\'{e}n wave has large current density along the wave front that can be dissipated to
   heat the coronal bottom.

   The concave structure of the transition region can also be important 
   because it produces the steep gradient of Alfv\'{e}n speed not only in 
   vertical direction but also in horizontal direction.
   The strong inhomogeneity in horizontal direction can proceed the so called phase mixing of Alfv\'{e}n waves
   \citep{1983A&A...117..220H}.
   We think the phase mixing just above the transition region stimulates the coronal heating
   in our simulation.

   The acoustic heating rate is also important at $r-r_{\rm s}\approx0.04$ $r_{\rm s}$ (30 Mm)
   The acoustic heating originates from the shock waves that are formed in the chromosphere 
   and injected to the coronal bottom \citep{1982ApJ...254..806H}.


   \subsection{Heating in the solar wind acceleration region}

\begin{figure}
     \begin{center}
      \includegraphics[scale=1.0]{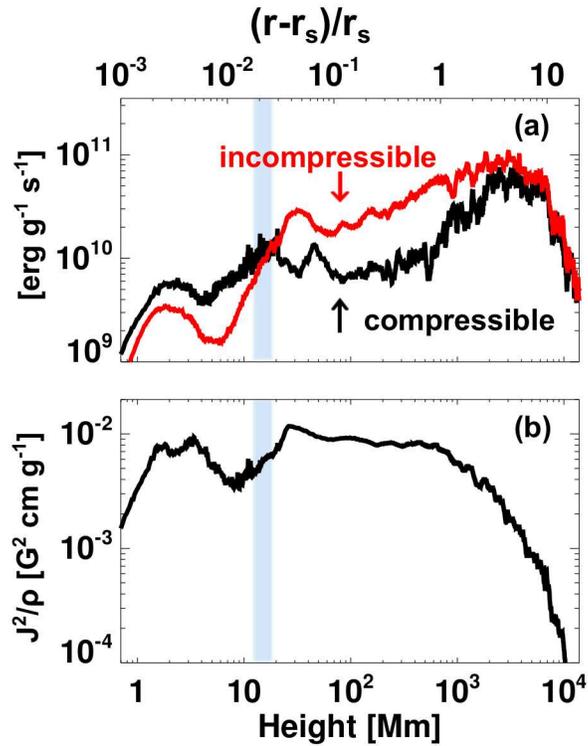}
         \caption{
                 Comparison between compressible and incompressible components of magnetic heating rate 
                 as a function of height (a), and mean profile of current density per unit mass (b).
                 The black/red solid line in panel 
                 (a) corresponds to compressible/incompressible heating rate.
                 The blue shaded area corresponds to the transition region.
                 }
         \label{fig:cmp_icmp_compare}
     \end{center}
   \end{figure}

   In the solar wind acceleration region ($r-r_{\rm s}> r_{\rm s}$ or 7$\times$10$^2$ Mm), 
   incompressible heating mechanisms are dominant.
   The incompressible heating mechanism here means the mechanisms such as MHD turbulence or phase mixing.
   Figure \ref{fig:cmp_icmp_compare}a shows that the incompressible component dominates the compressible component.
   The compressible component is derived in the way that we sum up $Q_{a}$ over the region
   where the velocity divergence is negative.
   The rest of $Q_{a}$ is referred to as the incompressible component.
   The incompressible component exceeds the compressible component above $r-r_{\rm s}=0.03$ $r_{\rm s}$ (21 Mm).
   Therefore the plasma heating in the solar wind are dissipation of magnetic energy through incompressible process, 
   although the heating is done through numerical dissipation.

   Extension to 3D could be essential in terms of the property of MHD turbulence in the solar wind.
   Reduced MHD that excludes the compressibility from the usual MHD is
   useful approach for MHD turbulence in the solar wind and
   extensive studies have been done in the reduced MHD formulation so far
   \citep{1999ApJ...523L..93M,2003ApJ...597.1097D,2007ApJ...662..669V,2011ApJ...736....3V}.
   Although the compressibility turns out not to contribute the total heating rate directly, 
   it produces large density fluctuations that could change the reflection rate of Alfv\'{e}n waves.
   Since the efficiency of MHD turbulence depends on the amount of the reflection waves,
   the compressibility could affect the total heating rate indirectly.


\section{Discussions} \label{sec:discussion}

   Thanks to the new analysis method described in section \ref{sec:data_analysis} and appendix \ref{app:method}, 
   we can derive more accurate profile of the heating rate than that was reported by MS12.
   Although the numerical simulation shown here is almost the same as that of MS12,
   the new analysis method and higher numerical resolution lead us two different interpretations about 
   heating mechanisms in the corona.
   
   First, MS12 suggested that the shock dissipation and MHD turbulence are 
   the main heating mechanisms at the coronal bottom.
   We concluded that, instead of MHD turbulence, Alfv\'{e}n waves with curved wedge shape
   originate from the chromospheric fast shocks is main heating mechanisms at the coronal bottom.
   Phase mixing due to the concave transition region may also stimulate the coronal heating.
   The heating rate derived in this paper is more reliable than that of MS12.
   Next, MS12 suggested that MHD turbulence is
   the dominant heating mechanisms in the solar wind.
   The finer radial resolution reveals that
   we do not achieve the shallower power spectrum like Kolmogorov-type turbulence.
   We confirm that the magnetic energy are converted to the thermal energy through incompressible processes,
   although it is difficult to distinguish what kind of heating mechanisms operate in the solar wind.

   We derive heating rate in our numerical simulation that inevitably has numerical dissipation larger than
   that of the realistic solar atmosphere.
   Therefore we should discuss the validity or the robustness of the heating mechanisms found in our simulations, 
   although we do need further numerical experiments to investigate whether our results depend on numerical resolution.
   Shock heating rate would not depend on the grid size and microscopic physics like 
   viscosity or resistivity as far as Rankine-Hugoniot relations are satisfied.
   Since our numerical scheme (HLLD scheme) is conservative and meditated to capture MHD shocks,
   the shock heating in our numerical simulation should not be just a numerical heating but a physical one.

   Heating events at the coronal bottom are associated with curved wedge shape structure described in section 
   \ref{sec:heating}. This structure would be Alfv\'{e}n wave front with large current density that was 
   stored in the chromospheric fast shocks. Whether this structure keeps large current density or not 
   might depend on the numerical resolution. The dynamic simulations with high resolution are 
   needed to justify this heating mechanism.

   The heating by phase mixing depends on viscosity or resistivity so that 
   our results could be affected by numerical resolution.
   The phase mixing in our simulation turns on because of the horizontal phase speed difference that 
   is created by the collision between the transition region and shock waves.
   Then the duration that phase mixing operates is finite and determined by the Alfv\'{e}n crossing time
   between peaks and troughs of the transition region.
   The finite duration for phase mixing results in the finite thickness of the current density.
   Therefore this will prevent phase mixing from cascading 
   to the dissipation scale that is determined by microscopic physics,
   although, in our simulation, Alfv\'{e}n waves dissipate due to insufficient grid numbers in $\phi$ direction.
   Therefore there is a possibility that the phase mixing discussed here may be weaken 
   to operate if we increase grid numbers.
   Extension to 3 dimensional simulation could stimulate phase mixing by increasing the effective viscosity 
   or resistivity through instability due to velocity or magnetic field shear.
   Or inhomogeneity in $\theta$ direction makes phase mixing less effective as was suggested by 
   \cite{1991ApJ...376..355P}.

   In the present study, we only consider shear Alfv\'{e}n fluctuations ($V_{\theta}$) 
   as the energy injection from the photosphere.
   However the realistic fluctuations would also be associated with kink and 
   torsional types in the complex solar atmosphere without translational symmetry ($\partial _\theta = 0$).
   Torsional Alfv\'{e}n waves could be generated by the vortex flows observed in the photosphere
   \citep{1988Natur.335..238B}, and the connection between the vortex flows and 
   swirling motions in the upper atmosphere are observed recently \citep{2012Natur.486..505W}.
   Since the difference in nonlinear behaviors between torsional and shear mode is
   discussed by \cite{2011A&A...526A..80V}, 
   dynamic behaviors could be affected if we include torsional Alfv\'{e}n waves.
   This aspect will be explored in future work by using axisymetric 2.5 dimensional simulations or 
   full 3 dimensional simulations.

   Since we only prescribe Alfv\'{e}nic fluctuations that cannot be linearly converted into compressible mode, 
   the nonlinear development of Alfv\'{e}n waves is important to heat the atmosphere.
   The wave nonlinearity $V_{\theta}/V_{\rm A}$ is proportional to $\rho^{1/4}B^{-1}$ if we consider the
   wave energy flux conservation in an static atmosphere, $\rho V_{\theta}^2 V_{\rm A} A=$ const, where $A$ is the cross section of the flux tube.
   This means that not only the density structure but also the expansion rate of the magnetic field are important 
   for the development of the wave nonlinearity.
   The previous 1 dimensional simulations artificially defined the cross section of flux tubes.
   The magnetic field in our model has nearly the potential field structure 
   similar to that in \cite{2003ApJ...599..626B}, which makes our simulation more realistic than 
   the 1 dimensional simulations.
   Our model, however, cannot describe the non-magnetic atmosphere below the magnetic canopy, 
   since potential field structure needs the surrounding magnetic pressure to support the strong magnetic flux tube.
   For more realistic simulation, we should implement the magnetohydrostatic atmosphere
   in the similar way \cite{2005ApJ...631.1270H} did to investigate the compressible wave propagation.

   From the chromospheric observations with HINODE/SOT, \cite{2007PASJ...59S.655D} suggested that 
   there are two types of spicules: 
   Type I spicules exhibit up-and-down motion and have relatively slow ascending speed while
   Type II spicules fade out and often have high speed upward motion larger than 100 km s$^{-1}$.
   If we regard the up-and-down motion of the transition region as spicules, 
   the radial velocity of our spicules is usually 60 km s$^{-1}$ at most.
   Therefore the collision between shock waves and the transition region could not
   lift up the plasma so rapidly as Type II spicules do and
   other mechanisms will be needed \citep[e.g.][]{2011ApJ...736....9M}.
   It should be noted however that the existence of the Type II spicules 
   is still under debate so far \citep{2012ApJ...750...16Z,2012ApJ...759...18P}.


\section{Summary} \label{sec:summary}

   We have implemented a 2.5 dimensional MHD simulation that resolves the propagation and the dissipation of 
   the Alfv\'{e}n waves in the solar atmosphere.
   The hot corona and the high speed solar wind are reproduced as a natural consequence of 
   the Alfv\'{e}n wave injection from the photosphere, as was shown in MS12.
   However, the detailed analysis of the heating rate leads us to the different interpletation as follows.
   \begin{itemize}
   \item Alfv\'{e}n waves with curved wedge shape generated by the chromospheric fast shocks heat 
         the coronal bottom while MS12 suggested heating by MHD turbulence.
   \item Anisotropic turbulent cascade that was found in the simulation of MS12 
         turned out to disappear when increasing radial resolution.
         The heating in the solar wind operates through the magnetic energy conversion with
         incompressible process, though we could not distinguish the specific heating mechanisms so far.
   \end{itemize}


\section*{Acknowledgments}

We thank the referee, professor P.Cargill for giving us fruitful comments on our manuscripts.
Numerical computations were carried out on Cray XT4 at Center for Computational Astrophysics, CfCA, 
of National Astronomical Observatory of Japan. 
The numerical calculations were also carried out on SR16000 at YITP in Kyoto University.
Takuma Matsumoto gratefully acknowledges the research support in the form of fellowship 
from the Japan Society for the Promotion of Science for Young Scientists.
This work was also supported in part by Grants-in-Aid for Scientific Research from the MEXT of Japan, 22864006.


\hyphenation{Post-Script Sprin-ger}

\appendix

   \section{Resolution Test} \label{sec:res_test}
   In order to check the dependence of our results on the numerical resolution,
   we have performed high resolution runs.
   Since we do not have enough CPU power, we will measure the energy flux
   at the corona resulting from a sinusoidal Alfv\'{e}nic fluctuation at the photosphere.
   For simplicity, we ignore the radiation cooling and thermal conduction.
   Initial magnetic field is the same as the one described in section \ref{sec:model}.
   As for temperature distribution, we set a tangent hyperbolic function,
   \begin{eqnarray}
     T = T_0 + {1\over 2}(T_c-T_0)(1+\tanh{{x-h\over w}}),
   \end{eqnarray}
   to mimic the photosphere, transition region, and corona, 
   where $T_0=6,000$ K, $T_c=2$ MK, $x$ is height, $h=3,000$ km, and $w=500$ km.
   The photospheric density is set to be $10^{-7}$ g cm$^{-3}$ and
   the hydrostatic equation is solved to get the density stratification.
   \begin{figure}
      \begin{center}
      \includegraphics[scale=1.0]{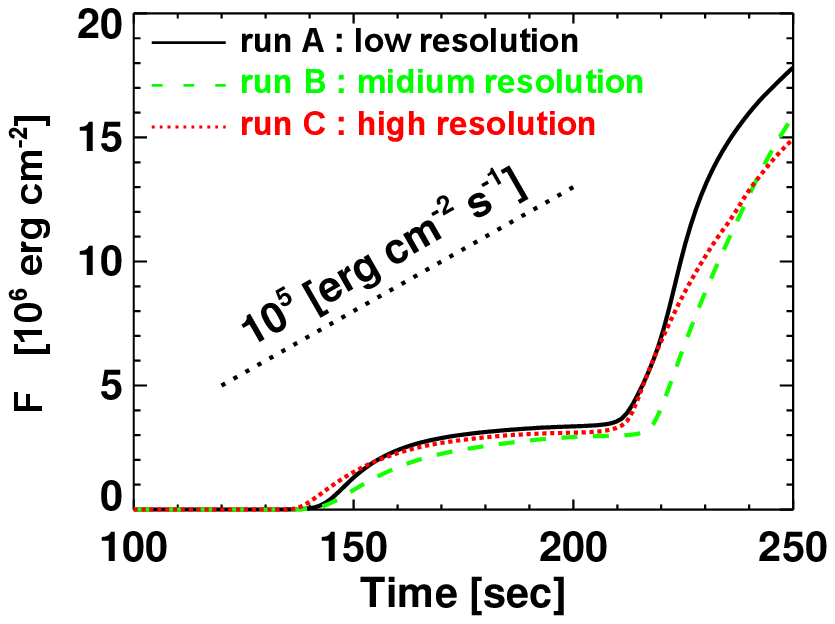}
         \caption{ Accumulated Poynting flux at 10 Mm (see eq \ref{eq:acc_flux} in text) as a function of time.
                 The black solid, green dashed, and red dotted lines represent the 
                 results from run A (low resolution), run B (medium resolution), and
                 run C (high resolution).
                 The slope of the dotted line corresponds to the energy flux of
                 10$^5$ erg cm$^{-2}$ s$^{-1}$.
                  }
         \label{fig:res_test}
      \end{center}
   \end{figure}
   Alfv\'{e}n wave is driven by specifying a sinusoidal velocity perturbation, 
   \begin{eqnarray}
     V_\theta = 2 \sin{{2\pi t\over180{\rm~sec} }} {\rm~km~s^{-1}}.
   \end{eqnarray}
   The velocity perturbation is set to be zero when $t>180$ sec.
   Three simulations have been performed, changing the grid number shown in table \ref{tab:run}.
   The mesh sizes are constant below 10 Mm.
   The vertical mesh sizes are enlarged above 10 Mm so that the maximum height is greater than 800 Mm.

   Figure \ref{fig:res_test} shows the Poynting flux averaged over $\phi$ and integrated over time at 10 Mm,
   \begin{eqnarray}
     F \equiv {1\over \Delta \phi}\int_{\Delta \phi} d\phi \int_0^t- B_\theta V_\theta B_r  dt',\label{eq:acc_flux}
   \end{eqnarray}
   where we just pick up the contribution from $\theta$ component that is larger than the other components.
   During $140<t<210$ sec, the first half of the injected sinusoidal wave passed,
   while the second half of the wave passed after $t=210$ sec.
   As long as the Poynting flux is concerned, the relative error in this simulation is 
   less than 50 \%.

   \begin{table}
     \begin{center}
       \begin{tabular}{lcc} \hline
         &grid number & grid size [km] $^*$ \\ \hline \hline
         run A & 2048x32& 25x93.8 \\
         run B & 2048x128& 25x23.4 \\
         run C & 4096x512& 6.1x5.9 \\ \hline
       \end{tabular}
     \end{center}
     \caption{$^*$ This grid size is measured below 10 Mm. Above 10Mm, 
             the vertical grid size is enlarged.
             }
     \label{tab:run}
   \end{table}

   \section{Estimation of heating rate} \label{app:method}
   In section \ref{sec:data_analysis}, we gave the basic idea to
   estimate the heating rate.
   We will describe the detailed numerical procedure in this appendix.
   The method can be applied to the numerical scheme that uses the finite volume method with
   some kind of Riemann solver flux estimation.
   Although the description here is for 2.5 dimensional spherical coordinate system,
   application for the other orthogonal coordinate system is straightforward.

   In the finite volume method, internal energy is updated after all the other conservative variables
   are updated. Then the discretized equation for internal energy update will be
   \begin{eqnarray}
     \Delta ur^2 = \Delta {\cal E} - \Delta { \rho V^2r^2 \over 2} - \Delta {B^2 r^2 \over 2} \label{eq:des_int},
   \end{eqnarray}
   where $\Delta f \equiv (f_{i,j}^{n+1}-f_{i,j}^{n})/\Delta t$ and $f$ is arbitrary variable, and
   ${\cal E} = r^2 E$.
   The superscript indicates a certain time step $n$ and the subscripts $i$ and $j$ represent the discretized index of the spatial coordinate $r$ and $\phi$, respectively.
   $\Delta t$ represents the time interval between the time step $n$ and $n+1$.

   The change rate of the kinetic energy can be decomposed into several parts within the rounding error as follows.
   \begin{eqnarray}
     \Delta { \rho V^2 r^2 \over 2}&=& {1\over \Delta t} \left[ { \left({\cal M}^{n+1}\right)^2 \over 2 {\cal R}^{n+1}} 
     - { \left({\cal M}^{n}\right)^2 \over 2 {\cal R}^{n}} \right] \nonumber \\
     &=& \sum_{\alpha = r,\theta,\phi} { {\cal M}^{n+1/2}_\alpha \Delta {\cal M}_\alpha \over {\cal R}^{n+1} }
     - { \left( {\cal M}^n_\alpha \right)^2 \Delta {\cal R} \over 2 {\cal R}^{n+1} {\cal R}^n},\label{eq:des_kin}
   \end{eqnarray}
   where $f^{n+1/2} \equiv (f^{n}+f^{n+1}) / 2$, 
   ${\cal R} = r^2 \rho$, and ${\cal M}_\alpha = r^2 \rho V_\alpha$.
   In equation (\ref{eq:des_int}) and (\ref{eq:des_kin}), $\Delta {\cal E}$ and $\Delta {\cal M}_\alpha$ can be
   described by the spatial difference of the numerical flux. If we extract the terms related to gas pressure 
   from the spatial difference, the equation (\ref{eq:des_int}) can be divided into two parts as was done in 
   section \ref{sec:data_analysis}, as follows.
   \begin{eqnarray}
     \Delta u = Q_{n} + Q_{a}, \label{eq:decomp_2d}
   \end{eqnarray}
   where
   \begin{eqnarray}
     Q_{n} = - { \Delta _r \left[ \left( u+P_g \right) V_r r^2 \right] \over r_i^2}  
            - { \Delta _\phi \left[ \left( u+P_g \right) V_\phi \right] \over r_i} 
            + {{\cal M}_{r;i,j}^{n+1/2} \over {\cal R}_{i,j}^{n+1} r_i^2} \Delta _r \left( P_g r^2 \right)
            + {{\cal M}_{\phi;i,j}^{n+1/2} \over {\cal R}_{i,j}^{n+1} r_i} \Delta _\phi \left( P_g \right),
   \end{eqnarray}
   where $\Delta_r f \equiv (f_{i+1/2,j}^*-f_{i-1/2,j}^*)/\Delta r_i$ and 
   $\Delta_\phi f \equiv (f_{i,j+1/2}^*-f_{i,j-1/2}^*)/\Delta \phi_j$.
   $Q_{n}$ can roughly be regarded as the sum of adiabatic heating and heating 
   at hydrodynamical shocks, while $Q_{a}$ consists of the rest of all the 
   entropy generation, the sum of numerical viscous dissipation by velocity 
   shear and numerical resistive dissipation of magnetic field. 
   Although $Q_{n}$ may be estimated by the central differences, we found that 
   the positivity of $Q_{a}$ is significantly improved when we use 
   the variables from Riemann solver for the inside of the numerical difference.
   We can not derive all the variables individually, since HLLD scheme is approximate Riemann solver and can not derive gas pressure self-consistently.
   Instead, we only know averaged value of $E,P_T,\mathbf{V},\mathbf{B}$, and $\rho$ at the cell surface in HLLD scheme.
   Therefore when we use HLLD scheme, we adopt further approximation as follows.
   \begin{eqnarray}
     P_g &\equiv& P_T - {B^2 \over 2} \label{eq:start_hlld}\\
     u+P_g &\equiv& E - {1\over2} \rho V^2 + P_T - B^2 \\
   \end{eqnarray}
   The second term in (\ref{eq:decomp_2d}), $Q_{a}$, can be derived if we subtract $Q_{n}$ from $\Delta u$.
   \subsection{dissipation of linear wave in MHD}
   In order to investigate the property of $Q_{n}$ and $Q_{a}$, we have performed 
   test simulations for dissipation of linear MHD wave.
   We use 2D ($x,z$) Cartesian grid 
   with grid numbers of ($N_x,N_z$) = (128, 64). The spatial domain is $(0<x<\sqrt{5})$ and 
   $(0<z<\sqrt{5}/2)$. The initial conditions are described as follows.
   \begin{eqnarray}
     \mathbf{q}^{\rm initial} &=& \bar{\mathbf{q}} + \epsilon \mathbf{r}_k \sin{2\pi x'}, \\
     x'  &=& x \cos{\theta} + z \sin{\theta},
   \end{eqnarray}
   where 
   \begin{eqnarray}
     \mathbf{q}&=& \left( \rho, V_{\rm x'}, V_{\rm y'}, V_{\rm z'}, B_{\rm x'}, B_{\rm y'}, B_{\rm z'}, P_g\right),
   \end{eqnarray}
   \begin{eqnarray}
     \bar{\mathbf{q}}&=& \left( 1, 0, 0, 0, B_0 \cos{\theta_{\rm B}}, 0, B_0\sin{\theta_{\rm B}}, 1/\gamma \right),
     \label{eq:background}
   \end{eqnarray}
   and $\mathbf{r}_k$ is the right eigen vectors that are described in subsection {\ref{sec:eigen_vector}}.
   The wave amplitude, $\epsilon$, is set to be $10^{-2}$ and $\theta$ is the angle of the wave number vector
   measured from $x$ axis where $\tan^{-1} \theta=2$.
   $\gamma (=5/3)$ is specific heat ratio and $\theta _{\rm B}$ is the angle between the wave number vector and
   the background magnetic field.
   Periodic boundary conditions are posed on both boundaries.
   Although we do not include any explicit dissipation term, total kinetic and magnetic energy 
   will be decreasing by numerical dissipation. 
   We have performed several simulations with different plasma beta and $\theta_{\rm B}$.

   \begin{figure*}
      \includegraphics[scale=1.0]{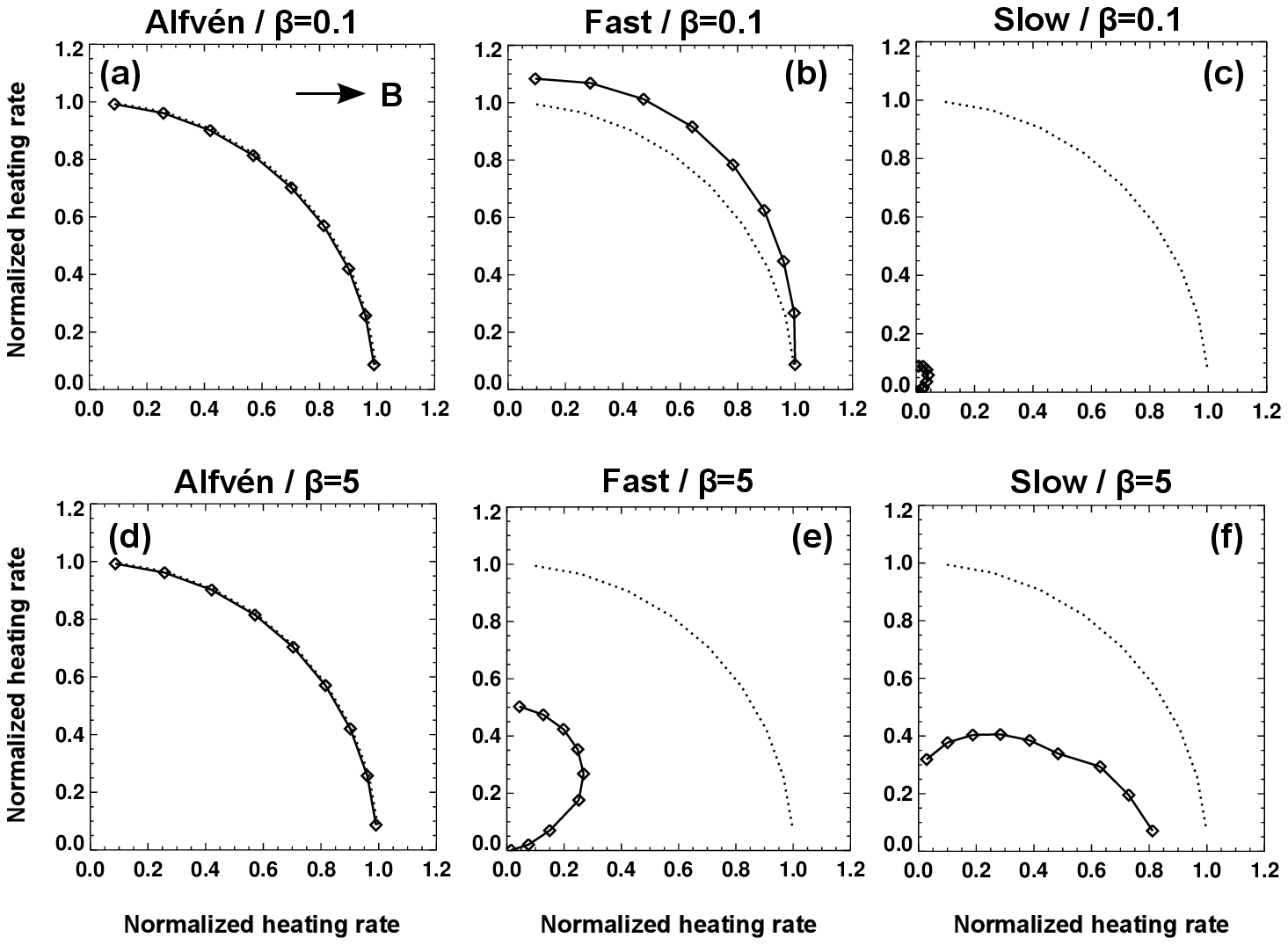}
         \caption{Normalized numerical heating rate in linear MHD wave propagation with different
                  plasma beta and $\theta_{\rm B}$ .
                  }
         \label{fig:dissip_test}
   \end{figure*}

   For each run, we can measure the energy loss rate of kinetic and magnetic energy 
   averaged over space and time.
   In the meantime, we can obtain $Q_{a}$ averaged over space and time.
   The averaging procedures are done over all the spatial domain and over one wave period.
   Then we can obtain $Q_{a}$ normalized by the energy loss rate for each run.
   Figure \ref{fig:dissip_test}a shows the normalized heating rate as a function of $\theta_{\rm B}$
   for Alfv\'{e}n wave with plasma beta equals to $0.1$.
   Each diamond corresponds to a single run, where the distance between the origin $(O)$ and the diamond $(P)$
   indicates normalized heating rate and the angle between the line $\overline{OP}$ and horizontal axis
   represents $\theta_{\rm B}$.
   We plotted a unit circle by the dotted line as a reference.
   Figure \ref{fig:dissip_test}a suggests that $Q_{a}$ is excellent indicator of numerical dissipation rate for 
   Alfv\'{e}n wave.
   The panels b-f of figure \ref{fig:dissip_test} corresponds to the runs for
   fast mode with $\beta=0.1$ (b), slow mode with $\beta=0.1$ (c),
   Alfv\'{e}n mode with $\beta=5$ (d), fast mode with $\beta=5$ (c), slow mode with $\beta=5$ (f).
   This figure suggests that $Q_{a}$ always gives good estimation for Alfv\'{e}n waves
   while $Q_{a}$ is good indicator for fast waves only in low beta plasma.
   For slow waves, $Q_{a}$ always underestimates numerical dissipation and 
   the dissipation mainly originates from $Q_{n}$.

   These results suggest that $Q_{a}$ and $Q_{n}$ corresponds to the
   dissipation rates of magnetic and gaseous energy, respectively.
   Since the fast waves in high(low) beta plasma have large(small) thermal energy,
   $Q_{a}$($Q_{n}$) becomes the dominant dissipation term.
   On the other hand the slow waves in high(low) beta plasma have small(large) thermal energy 
   so that $Q_{a}$($Q_{n}$) has considerable effects on total dissipation.
   

   \subsection{eigen vectors}{\label{sec:eigen_vector}}
   Here we show the right eigen vectors used in the test simulations for linear MHD waves.
   The right eigen vectors under the background condition
   of eq \ref{eq:background} can be written as
   \begin{eqnarray}
     {\mathbf{r}}_{\rm f} = \left(
     \begin{array}{c}
       \alpha_{\rm f}\\
       \alpha_{\rm f} c_{\rm f} \\
       0 \\
       -\alpha_{\rm s} c_{\rm s}\\
       0\\
       \alpha_{\rm s} \\
       \alpha_{\rm f}
     \end{array}     \right) ,
   \end{eqnarray}
   \begin{eqnarray}
     {\mathbf{r}}_{\rm s} = \left(
     \begin{array}{c}
       \alpha_{\rm s} \\
       \alpha_{\rm s} c_{\rm s} \\
       0 \\
       \alpha_{\rm f} c_{\rm f}\\
       0\\
       -\alpha_{\rm f} \\
       \alpha_{\rm s} 
     \end{array}     \right) ,
   \end{eqnarray}
   \begin{eqnarray}
     {\mathbf{r}}_{\rm A} = \left(
     \begin{array}{c}
       0 \\
       0 \\
       1 \\
       0 \\
       -1 \\
       0 \\
       0
     \end{array}     \right),
   \end{eqnarray}
   where $\mathbf{r}_{\rm f}$, $\mathbf{r}_{\rm s}$, and $\mathbf{r}_{\rm A}$ represent
   the right eigen vectors of fast, slow, and Alfv\'{e}n wave, respectively and
   \begin{eqnarray}
     c_{\rm f,s}^2= {1\over2} \left[ 1 + V_{\rm A}^2 \pm \sqrt{\left( 1+V_{\rm A}^2\right)^2 - 4 V_{\rm A}^2 \cos^2{\theta_{\rm B}}} \right],
   \end{eqnarray}
   \begin{eqnarray}
     \alpha_{\rm f}^2 &=& {1 - c_{\rm s}^2 \over c_{\rm f}^2 - c_{\rm s}^2}, \\
     \alpha_{\rm s}^2 &=& {c_{\rm f}^2 -1 \over c_{\rm f}^2 - c_{\rm s}^2},
   \end{eqnarray}
   where $\alpha_{\rm f/s}$ is defined as positive value.

\bsp

\label{lastpage}

\end{document}